\documentstyle[aps, eqsecnum]{revtex}
\begin{document}

\title{RELATIVITY WITHOUT RELATIVITY}

\author{Julian Barbour\cite{jb}}
\address{College Farm, South Newington, Banbury, Oxon, OX15 4JG, UK}
\author{Brendan Z. Foster\cite{bf}}
\address{Graduate School, Physics Department, Univ. of Maryland, College
Park, Maryland,  USA}
\author{Niall \'O Murchadha\cite{nom}}
\address{Physics Department, University College, Cork, Ireland}
\date{April 2nd, 2002}

\maketitle

\begin{abstract}
We give a derivation of general relativity (GR) and the gauge principle
 that is novel in presupposing neither spacetime nor the relativity
principle.  We consider a class of actions defined on superspace (the
 space of Riemannian 3-geometries on a given bare manifold). It has
two key properties. The first is symmetry under 3-diffeomorphisms. This
is the only postulated  symmetry, and it leads to a constraint linear
in the canonical momenta. The second property is that the Lagrangian
is constructed from a `local' square root of an expression quadratic
in  the velocities. The square root is `local' because it is taken
before integration over 3-space. It gives rise to quadratic
constraints that do not correspond to any symmetry and are not,
in general, propagated by the Euler--Lagrange equations. Therefore
these actions are internally  inconsistent.  However, one
action of this form is well behaved: the Baierlein--Sharp--Wheeler (BSW
\cite{bsw}) reparametrisation-invariant action for GR. From
this viewpoint, spacetime symmetry is emergent. It appears as a
`hidden' symmetry in the (underdetermined) solutions of the
Euler-Lagrange equations, without being manifestly coded into the action
itself.
In addition, propagation of the linear diffeomorphism constraint
together with the quadratic square-root constraint acts as a striking
selection mechanism beyond pure gravity. If a scalar field is included
in the configuration space, it must have the
same characteristic speed as gravity. Thus Einstein causality emerges.
Finally, self-consistency requires that any 3-vector field must
satisfy Einstein causality, the equivalence principle and, in
addition, the Gauss constraint. Therefore we recover the standard
(massless) Maxwell equations.

\end{abstract}

\section{Introduction}

Traditionally the 3+1 dynamical formulation of GR, with its
associated Hamiltonian and Lagrangian structures, is obtained by
projection from spacetime. Geometrodynamics has a
four-dimensional symmetry from the start. In contrast, the
classic study of Hojman, Kucha\v r, and Teitelboim \cite{hkt1}
starts with a three-dimensional fully constrained Hamiltonian
and requires its constraint algebra to close by reproducing the
standard Dirac algebra for GR. In this way, they recover the ADM
Hamiltonian. However, by insisting on the specific structure
`constants' of the Dirac algebra, they still presuppose
spacetime. In fact, this is unnecessary. We shall show that mere
closure of the algebra is enough to obtain GR.

We shall begin by explaining how the Baierlein--Sharp--Wheeler
action, which is at the centre of our investigation, arose from
the ADM formalism \cite{adm}. In the ADM approach, one starts
from spacetime (a pseudo-Riemannian 4-manifold). One observes
that the `surfaces' of constant label time are spacelike
(Riemannian) 3-geometries. Geometrodynamics is the evolution of
these 3-geometries. Choosing 3-coordinates, one gets a smooth
curve of 3-metrics, $g_{ij}(x^i, t)$, $i,j = 1,2,3$ with $x^i$
the coordinates. The $g_{ij}$'s depend on the slice (labelled by
$t$), on the coordinates and on the point on the slice. The
coordinates can be transformed because only the 3-geometry
matters, not the 3-metric. Moreover, the transformation can be
changed freely from slice to slice as well as from point to
point.

The key concepts in the ADM formalism are the 3-metrics $g_{ij}$,
the lapse $N$ and the shift $N^{i}$. The lapse measures the rate
of change of proper time w.r.t. the label time, while the shift
determines how the coordinates are laid down on the successive
3-geometries. Prior to the transition to the Hamiltonian, the
standard Hilbert--Einstein action for matter-free GR is rewritten,
after divergence terms have been omitted,
in the 3+1 form \begin{equation}
I= \int\sqrt{g}N\left[R + K^{ij}K_{ij} - {\rm
tr}K^2\right]d^3x.
\label{first}
\end{equation}

Here $R$ is the three-dimensional scalar curvature, and $K_{ij} =
-(1/2N)(\partial g_{ij}/
\partial t - N_{i;j} - N_{j;i})$ is the extrinsic curvature. From here
 the transition made by BSW
\cite{bsw} is trivial. They first replaced $K_{ij}$ in the action
by $k_{ij} = \partial g_{ij}/
\partial t - N_{i;j} - N_{j;i}$, the unnormalised normal
derivative, to give
\begin{equation}
I= \int\sqrt{g}\left[NR + {1 \over 4N}\left(k^{ij}k_{ij} - {\rm
tr}k^2\right)\right]d^3x.
\label{second}
\end{equation}
They varied this action with respect to the lapse and found an
algebraic expression for it,
 \begin{equation} N =
\sqrt{{k^{ij}k_{ij} - {\rm tr}k^2 \over 4R}}. \label{third}
\end{equation}
This, of course, is clearly consistent with the ADM Hamiltonian
constraint. In turn, this expression for $N$ is substituted back
into Eq.(\ref{first}) to obtain the BSW Lagrangian
\begin{equation} I = \int\sqrt{g}\sqrt{R}\sqrt{k^{ij}k_{ij} -
{\rm tr}k^2}. \label{fourth} \end{equation}

There is a large class of four-dimensionally generally covariant
theories of matter fields coupled to gravity in which the
Lagrangian is a sum of potential-type terms (like $R$) +
 quadratic products of the `proper' time derivatives, all multiplied
by $N$. In all of them, one can repeat the process above,
solve for the lapse algebraically and pass trivially to the BSW form.

The BSW action has attracted relatively little attention. It was
initially proposed, especially by Wheeler, as the
starting point of a method of solving the initial-value problem of
GR. The original BSW paper \cite{bsw} formulates the
`thin-sandwich problem', which will be mentioned later. However,
in this paper, we propose to take (\ref{fourth}) as paradigmatic
for the dynamics of the universe. We shall recall some basic
concepts and identify the two key properties of BSW-type actions. Then, in
view of its relative unfamiliarity, we
shall give a fairly extended outline of our approach, which we
call {\it{the 3-space approach}}.

Wheeler \cite{jaw} called the set of Riemannian 3-geometries on a
given topology {\it{superspace}}, which is the ADM configuration
space and the space we shall use here. However, we actually
believe that {\it{conformal superspace}}, which is obtained by
quotienting superspace by three-dimensional conformal
transformations, may well be the true configuration space for
gravity
\cite{bom}, but in this paper we work in superspace. The
fundamental object we consider is a `curve in superspace', a
sequence of 3-geometries, labelled by some parameter $\lambda$. We
use $\lambda$ rather than $t$ because in our approach time emerges
from geodesic-type curves on configuration spaces. This timeless
approach is motivated below and in \cite{bb}.

For well-known technical reasons, we cannot work directly with
superspace; to integrate and perform other key mathematical
operations, we must introduce coordinates. Therefore the actual
objects we consider are curves of 3-metrics $g_{ij}(x^i,
\lambda)$, for which we define an action $A$. It must not depend
on the choice of coordinates, i.e., $A$ must be truly a function
on curves in superspace rather than just on curves in {\it Riem},
the space of 3-metrics. This leads us to the first key property of
the BSW-type actions.

We know that $A$ will depend on $g_{ij}$, its spatial
derivatives, and also on its `velocity' $\partial
g_{ij}/\partial\lambda$. Moreover, $A$ must be coordinate
invariant. In the case of $\lambda$-independent transformations,
all we need ensure is that the integrand of $A$ is a 3-scalar,
i.e., that we use covariant, rather than ordinary, derivatives and
that the indices upstairs and downstairs match. However,
crucially, we must also consider $\lambda$-dependent
transformations, i.e, ones that differ from slice to slice. They
have the form $x^i \rightarrow x'^{i}(x^j, \lambda)$ with inverse
$x^j(x'^i, \lambda)$. We now find
$${\partial g_{ij}' \over \partial \lambda} = {\partial x^a \over \partial
x'^{i}} {\partial x^b \over \partial x'^{j}}\left({ \partial
g_{ab} \over
\partial \lambda} - \xi_{a;b} - \xi_{b;a}\right)$$
where $\xi^a = -\partial x^a /\partial\lambda$, so the $\lambda$
derivative does not transform as a 3-tensor. This is a general
problem that arises whenever the `points' of a configuration space
have an internal (gauge) symmetry. Indeed, it will be argued in
\cite{barb} that this is the defining characteristic of
any gauge theory. In the present case, the `gauge' corresponds to
coordinate transformations on space. We overcome this problem by
having the metric velocity always appear as part of the combination $
k_{ab} = \partial g_{ab}/ \partial \lambda - \xi_{a;b} - \xi_{b;a}$,
where we regard $\xi^i$ as an arbitrary 3-vector field and minimize
the action with respect to it as well as with respect to the metric.
This process of adjusting for the gauge freedom is universal and is
appropriately called {\it{best matching}}. It was introduced to
implement the conviction of Leibniz, Mach and many others that motion
is relative \cite{bb}. We will write $\xi_{a;b} +\xi_{b;a} =
(K\xi)_{ab}$, because this expression is the Killing form of the
vector $\xi^i$. It is also the Lie derivative of $g_{ij}$ along
$\xi^i$. A further point is that we are constructing geodesic-type
actions, which requires $A$ to be reparametrisation invariant (see
below). This will be guaranteed because $\xi^{i}$ is actually a
velocity, so that the combination $ k_{ij} =
\partial g_{ab}/
\partial \lambda - \xi_{a;b} - \xi_{b;a}$ is linear in $1/d\lambda$.

Initially, we consider actions which depend only on the metric and
its velocity (and the vector $\xi$). We will later extend the
configuration space to include a scalar function $\phi(x^i,
\lambda)$ and a 3-vector field $A^i(x^i, \lambda)$. The velocities
of these fields will also include best-matching corrections. Since
they are based on the Lie derivative, all such correction terms
have a universal nature. However, their universal common origin
manifests itself in remarkably different ways because each
different bosonic field has its own different Lie derivative.
Eqns. (1.1)--(1.4) together with the expression for the
extrinsic curvature show how best matching is implemented in GR
through the Lie derivative. We shall show how this fact coupled with
the second key property of BSW-type actions leads to the novel
derivation of GR.

We now turn to this second key property of BSW actions, the
square roots, which make them similar to the action of Jacobi's
principle of classical mechanics. In fact, as Lanczos notes in his
beautiful book on the variational principles of mechanics
\cite{lanczos}, Jacobi-type square-root actions were
effectively used by Euler and Lagrange in their application of
Maupertuis's principle. However, it was Jacobi who clarified the
mathematics and cleanly separated two different principles. The first
is Hamilton's principle based on the familiar $T-V$ Lagrangian. It
gives dynamical curves in configuration space together with the speed
at which they are traversed with respect to an independent external
time variable. The second is the principle now known as {\it{Jacobi's
principle}}, which merely gives the dynamical orbit in configuration
space independently of the speed at which it is traversed. It is, in
fact, a geodesic principle based on the square root of a form
quadratic in the particle displacements. In the Kepler problem, it
yields the elliptical planetary orbits. We shall be interested in
their generalization to general dynamical systems.

It is important to distinguish between {\it{dynamical orbits}}
determined by a Jacobi-type principle and {\it{group orbits}},
generated in the configuration space by the action of a gauge
group. For us, the most important example of a group orbit arises
from coordinate transformations on our 3-spaces. These correspond
to 3-diffeomorphisms. All 3-metrics in {\it Riem} that can be
carried into each other by 3-diffeomorphisms lie on one orbit of the
three-dimensional diffeomorphism group. Since the notion of group
orbits is well known, it is enough to mention it.

However, it is necessary to say why Jacobi-type actions are needed
in the attempt to treat the universe as an {\it{isolated dynamical
system}}, as we do here. Newton took time as an independent existent.
But it is read from clocks and ultimately cannot be anything but a
clock reading. Clocks, in their turn, are simply mechanisms that
participate in the universe's dynamics. The theory of time and
clocks must be derived from the dynamics of the universe
\cite{bb}. The fact is that speed is not obtained by dividing an
infinitesimal displacement by an infinitesimal time but by another
displacement. This was manifest, even if easily forgotten, when
the rotation angle of the earth was the measure of time. A
moment's reflection should persuade the reader that true dynamics
establishes correlations between displacements, not between
displacements and time. Now a curve in a configuration space is
the complete expression of evolving correlations between
displacements. Therefore, universal dynamics is the study of such
curves. Moreover, the simplest such curves, i.e., the ones
determined by the fewest data, are geodesics. This is why we study
geodesics and Jacobi actions. For more details, see \cite{bb}.

In Sec. 2, we present Jacobi's principle for point particles. The
Jacobi action is the product of the square root of a potential
term multiplied by the square root of a kinetic term. We shall
discuss two different Jacobi-type actions. One is the standard,
`good', one; the other is a nonstandard, `bad', one. In the first,
the quadratic kinetic terms of the particles are summed, and then
we take the square root of the sum. In the second, we take the root
before we sum. We have found, to our surprise, that the `bad'
choice, which is realized in the BSW action (\ref{fourth}), is a
good idea. It leads to a massively over-constrained action but,
almost by magic, selects GR, in the BSW form, as the only nontrivial
case that works. We should like to mention here that, so far as we
know, the first person to realise that BSW both was a Jacobi-type
action and that it had a local square root and thus was not of the
standard form was Karel Kucha\v r. He pointed this out to one of us
(JB) in 1980 (see e.g.
\cite{bb}). Karel's observation, which was followed over the
intervening years by innumerable discussions with JB, was decisive
for the eventual appearance of this paper.

Thus, we take a reparametrisation-invariant and
coordinate-invariant Jacobi action of the form
\begin{equation}
A = \int d\lambda \int \sqrt{g} \sqrt{P} \sqrt{{T}} d^3x.
\label{1.1}
\end{equation}
Formally, this is defined on the Cartesian product of {\it Riem}
(the space of three-metrics) and the space of the fields $\xi$
considered earlier. However, the intention is to turn it into an
action defined on superspace by varying w.r.t. $\xi$. The parameter
$\lambda$ labels the `points' on trial curves in {\it Riem}, $g$ is
the determinant of
$g_{ij}$, the `potential' $P$ is a 3-scalar constructed from
$g_{ij}$ and its spatial derivatives, and $T$ is a kinetic term.

We assume that $T$ is quadratic in the `corrected metric
velocities' $k_{ij}$. Then either the `good' or the `bad' square
root makes the Lagrangian linear in $1/d\lambda$ and
reparametrisation invariant because of the $d\lambda$ outside.
Dirac \cite{dirac} noted that any action that is homogeneous of
degree one in the velocities will have canonical momenta that are
homogeneous of degree zero. They must then necessarily satisfy a
constraint that is primary, i.e., follows without any variation
already from the mere form of the momenta. Now Noether's theorem
(Part 2) tells us that reparametrisation invariance, which for
general actions of the form (\ref{1.1}) just allows a uniform
scaling of the velocities over the entire manifold, will yield one
constraint. In contrast, the `bad' actions of the form (\ref{1.1})
generate an independent constraint at each space point. This
mismatch between the constraints and the symmetry will be crucial.

We now seek stationary points of the action $A$ (\ref{1.1}) as a
functional of $g_{ij}$ and the gauge field $\xi^i$. We first
compute $p^{ij}$, the momentum conjugate to $g_{ij}$. This is the
variation of $A$ w.r.t. $\partial g_{ij}/ \partial \lambda$ which,
in turn, is identical to the variation w.r.t. $k_{ij}$. The
variation of $A$ w.r.t. $\xi$ is the variation w.r.t. $k_{ij}$
(which is just $p^{ij}$), times the variation of $k_{ij}$ w.r.t.
$\xi$, which is just the gradient of the variation of $\xi$. We
integrate once and get the standard ADM diffeomorphism constraint,
$p^{ij}_{~~;i} = 0$. This result has nothing to do with the square
root or any other structure of the action except the use of `best
matching' to implement diffeomorphism invariance. As we add extra
fields, we will get extra velocities, but each will have the
appropriate Lie-derivative correction with respect to the same
vector $\xi$. The striking universality of the results obtained
later arise because all fields in nature are subject to the same
diffeomorphisms generated by the one single field $\xi$.

Variation of the terms that involve the extra non-metric fields
will generate source terms in the diffeomorphism constraint, but
the universality of the Lie-derivative prescription ensures that
their form will be independent of the Lagrangian's detailed
structure. As a result, the diffeomorphism constraint seems to be
kinematic in origin, reflecting merely the 3-tensorial nature of
the considered fields. This is wrong. The diffeomorphism
constraint reflects one of the two key properties of the dynamics,
namely, that its purpose is to make the action depend solely on
the group orbits that constitute the points of superspace.

Because the coordinates are more or less freely specifiable at
each point of the 3-manifold, the resulting symmetry (an arbitrary
3-vector at each point) matches the constraint (the vanishing of a
3-tensor divergence at each point). Therefore, we have no reason
to expect that the evolution will fail to propagate this
constraint.

In the terminology of Bergmann adopted by Dirac \cite{dirac}, the
diffeomorphism constraint is a secondary constraint because it
arises from variation (the associated primary constraint is the
identical vanishing of the momentum conjugate to $\xi$ because $A$
does not depend on $\partial \xi/\partial \lambda$). However, as
noted above, there is already a nontrivial primary constraint in
the system. It follows from the mere definition of $p^{ij}$ that a
quadratic expression of the same basic form as the Hamiltonian
constraint of GR vanishes identically at each point of the
3-manifold. However, if we put the square root needed to construct
a geodesic-type action outside the integral sign, we get an
integral rather than a point constraint. Then the single resulting
constraint per manifold matches the one degree of freedom per
manifold due to the reparametrisation invariance. There is no
longer an independent constraint at each point of the manifold.
This is why we say the `bad' action (\ref{1.1}) is
over-constrained, and why it is no surprise that, in general, it
does not generate sensible dynamics. We call the Hamiltonian-type
constraint the {\it{square-root constraint}} because it arises
from the placing of the square root.

Our initial motivation for studying the actions (\ref{1.1}) was
because the BSW \cite{bsw} action for GR has this form. In Secs.
IV and V we give a fairly complete account of the BSW action,
especially the propagation of its constraints. We then make the
Legendre transformation from velocities to momenta and rediscover
the ADM Hamiltonian with its well-known complete freedom to `push
forward the evolution of spacetime' by different amounts across
the instantaneous spacelike 3-manifold. This remarkable property
of GR, which goes by various names including foliation invariance
and `many-fingered time', is thus the `hidden symmetry' of the BSW
action. It shows that an over-constrained system can nevertheless
be viable.

Section VI contains our first new result. We consider a general
action of the form (\ref{1.1}) on superspace. Specifically, we
assume: 1) $T$ is the most general quadratic kinetic form allowed
by a natural locality requirement; and 2) the potential term $P$ is
an arbitrary scalar function of $g_{ij}$ and its spatial
derivatives. We check the propagation of the constraints by the
equations of motion and find that in general the square-root
constraint is not propagated. We need a `hidden symmetry' in the
Euler--Lagrange equations to match the extra constraints on the
initial data. From this over-constrained perspective, it is
remarkable that nevertheless there exists a handful of actions
(\ref{1.1}) that do propagate the constraints. In them, $T$ must
have the standard DeWitt form (a result already obtained some years
ago by Giulini
\cite{g}) and $P = DR + \Lambda$, where $D$ and $\Lambda$ are
constants and $R$ is the scalar 3-curvature. By an overall
scaling, we can set $D = -1, 0$, or +1. The cases $D = + 1$ and $D
= -1$ give Lorentzian and Euclidean GR, respectively, both with
$\Lambda$ as a possible cosmological constant. The $D = 0$ case is
called {\it{strong coupling}} or {\it{strong gravity}} and is the
limit of standard GR for an infinite gravitational constant
\cite{sc}. One can further argue that the $D = 1$ case is the
only one for which the equations of motion can be written in
hyperbolic form and thus is the only viable candidate for a
dynamical system. All other actions force the 3-metric to be
flat and its velocity to vanish. Unlike the square-root constraint,
the diffeomorphism constraint is always propagated, as expected.

In Sec. VII, we see if the kinetic and potential terms for a
scalar field $\phi$ can be added to BSW gravity in the Lorentzian
case, which has an emergent `light cone'. Best matching uniquely
determines the kinetic term, but there is wide freedom in the
potential, for which we make the ansatz
\begin{equation}
 R \rightarrow  R   - {C\over
4}g^{ab}\phi_{,a}\phi_{,b} +\sum_n A_n \phi^n. \label{1.3}
\end{equation}

This is the standard (minimal-coupling) potential energy (with an
arbitrary constant $C$) and a polynomial with constants $A_n$. The
polynomial, which could in fact be replaced by an arbitrary function
of $\phi$ (but not its derivatives), gives no difficulty, but we must
have $C = 1$ for the constraint algebra to close. Therefore, the
canonical speed of
$\phi$ must match with the speed of the metric disturbances: the
scalar field must respect the gravitational light cone. This
result surprised us even more than the recovery of GR from pure
3-space dynamics, especially since the manner of its derivation
showed clearly that we had found a possible universal method of
generating the SR light cone. (Until we extend our approach to
include fermions, we cannot claim that
3-diffeomorphism invariance coupled with the local square-root is
{\it{the}} origin of the universal light cone.)

The result we have just described shows that minimal coupling (and
with it Einsteinian free fall) arises naturally in the BSW
approach. In Sec. VII, we also discuss a general range of
`dilaton' theories and show that, in many cases, they can
be written in the local square root form. Brans--Dicke theory is a
special case of such theories. We express its `vacuum' form as a
  BSW-type action. We further show that we can self-consistently couple
in a massive scalar field in the Brans--Dicke frame. Therefore,
it seems that Brans--Dicke theory is consistent with our approach

Life becomes even more interesting when we try to couple in a
3-vector field, which we denote by $A_a$ (Sec. VIII). The kinetic
term is again uniquely fixed. We consider a fairly general
potential energy expression for $A_{a}$:
\begin{equation}    U_A =  C_1A_{a;b}A^{a;b} + C_2A_{a;b}A^{b;a} +
C_3A^a_{;a}A^b_{;b} + \sum_kB_k(A^aA_a)^k,
\end{equation}
where $C_i$ and $B_k$ are constants. We include, thus, all
possible terms quadratic in the first derivative plus a general
undifferentiated polynomial term. We find that, at this level, the
polynomial terms do not hinder propagation of the square-root
constraint but that we must have $C_1 = -1/4$, $C_2 = +1/4$, $C_3
= 0$. Therefore, the potential-energy term becomes -(curl$A)^2/4$.
This means that here too we recover the common light cone for the
vector field from our putative universal method.

But now there is an even more striking result. As we shall see,
the structure of the 3-vector field is significantly more
complicated than in the scalar case. It generates not only the
above three conditions on the three constants $C_{i}$, but also an
extra secondary constraint requiring the divergence of the vector
momentum to vanish, essentially the
Gauss constraint. This too must propagate, and we can easily show
it does so only when all the polynomial terms vanish. This means we
have derived standard Maxwell  electrodynamics.  We find this result
truly striking. It suggests that gravity, the light cone, and massless
electrodynamics all arise from the local square root and the action of
the three-dimensional Lie derivative.

In a companion paper \cite{BFOM} we show that the
3-metric field, the vector field and scalar fields can interact
among themselves only in the form of complex scalar fields with
U(1) gauge coupling to the vector field and both scalar fields
minimally coupled to gravity. Thus we obtain classical gauge
theory. A paper just completed by Edward Anderson and one of us
\cite{AB} shows that in the BSW framework 3-vector fields can
interact among themselves only as Yang--Mills fields minimally
coupled to gravity.

The 3-space approach uses nothing but manifold geometry, the
mathematics that arose when Euclid's fifth axiom was seen to be
independent and could be omitted. It uses topology and bosonic
fields: scalar, vector, and 3-metric. As geometrical objects,
these are on a par. But their individual properties lead to a
hierarchy. The 3-metric is first among equals. Through its
covariant derivative and determinant $g$, it creates tensor
calculus and the densities that permit integration and variation.
It creates the highroad of dynamics \cite{iw}. Moreover, the Lie
derivatives impose a chain that arises from the intricate core of
the metric Killing form: gravity, light cone, gauge theory. The
effectiveness of {\it{geometry}} in establishing this structure
and logic in {\it{dynamics}} is impressive. Galileo said: ``He who
attempts natural philosophy without geometry is lost." He meant
3-geometry. Was spacetime an accident?

\section{square root actions}
We first formulate Newtonian particle mechanics without time,
using Jacobi's principle. It describes the orbits of
conservative dynamical systems in configuration space without
reference to motion. Jacobi's principle is very well known to
$N$-body specialists, who geometrize motion in {\it{configuration
space}} by exploiting the Riemannian metric defined there by the
kinetic energy of conservative mechanical systems \cite{lanczos}.
However, it is virtually unknown among relativists, who use
{\it{spacetime}} to geometrize motion. We have here a `cultural
divide' that we hope the present paper, which suggests that the
configuration space is more fundamental than spacetime, will help to
overcome.

For $N$ particles, labelled by $(m) = 1, \dots , N$, of masses
$m_{(m)}$ moving in a potential $V(x^i_{(1)}, \dots , x^i_{(N)})$,
the Jacobi action is \cite{lanczos}
\begin{equation}
    I_\textrm{G} =    2\int\sqrt{E -
V}\sqrt{T} d\lambda, \label{2.1a}
\end{equation}
where the constant $E$ is the total energy, $V$ is the potential
energy, $\lambda$ labels the points on trial curves and
$$  {T} = \sum_{(m) = 1}^N {m_{(m)} \over
2}{dx_{(m)}^i\over d\lambda}{dx_{(m)i}\over d\lambda}$$ is the
kinetic energy but with $\lambda$ in place of Newtonian time. The
subscript G on $I_\textrm{G}$ distinguishes the `good' action from
the `bad' one introduced later. The action (\ref{2.1a}) is
timeless since the label $\lambda$ could be omitted and the mere
displacements $dx_{(m)i}$ employed, as is reflected in the
invariance of $I_\textrm{G}$ under the reparametrization
\begin{equation}
\lambda \rightarrow f(\lambda).
\end{equation}

Jacobi's principle describes all Newtonian motions of one $E$ as
geodesics on configuration space. Its square roots are
characteristic, indeed essential, and are central in the 3-space
approach. They fix the structure of the action's canonical
momenta, which are
\begin{equation}
    p^{i}_{(m)} = {\delta {\cal L} \over \delta\left({d x_{(m)i} \over
d \lambda}\right)} = m_{(m)}\sqrt{E - V \over {T}} {d x^i_{(m)}
\over d \lambda}.\label{2.3a}
\end{equation}

Because $T$ (quadratic in the velocities) occurs under the square
root in the denominator while the velocity occurs linearly in the
numerator, the canonical momenta are homogeneous of degree zero.
They resemble direction cosines, which, if squared and added, give
1. We have
$$ {p^{i}_{(m)}p_{(m)i} \over 2m_{(m)}} = {E - V \over {T}}{m_{(m)}
\over 2}{dx_{(m)}^i\over d\lambda}{dx_{(m)i}\over d\lambda}.$$
If we sum the kinetic energies, they give $T$, cancelling the
numerator's $T$. This gives the {\it reparametrisation}, or {\it
square-root, identity}
\begin{equation}
\sum_{(m) = 1}^N{p^{i}_{(m)}p_{(m)i} \over 2m_{(m)}} = {E - V
\over {T}} \times {T} = E - V . \label{2.4a}
\end{equation}
Equation (\ref{2.4a}) seems to express energy conservation, but it
is actually an identity true on all curves in the configuration
space and not merely on-shell. In the Hamiltonian formalism,
(\ref{2.4a}) becomes a quadratic {\it{constraint}}.

The Euler--Lagrange equations are
\begin{equation}
{d  p^{i}_{(m)} \over d \lambda} = {\delta {\cal L} \over \delta
x_{(m)i}} = -\sqrt{{T} \over E - V}d^i_{(m)}V ,
 \label{2.5a}
\end{equation}
where $\lambda$ is still arbitrary. The solutions of Eq.(\ref{2.5a})
are, as they must be, parametrised curves in configuration space.
Within generalized Hamiltonian dynamics
\cite{dirac}, there is no guarantee that the Euler--Lagrange
equations used to propagate the canonical momenta will do so in
such a way that the identity (\ref{2.4a}) will be maintained. This
is why (\ref{2.4a}), and any other identity like it, is to be
regarded as a constraint whose propagation must always be checked.
In the present case, the Euler--Lagrange equations do conserve the
constraint:
\begin{equation}
\sum_{(m) = 1}^N{p_{(m)i}\over m_{(m)}}{d   p^{i} \over d \lambda}
= -\sum_{(m) = 1}^N{p_{(m)i}\over m_{(m)}}\sqrt{{T} \over E -
V}d^i_{(m)}V  = -\sum_{(m) = 1}^N\sqrt{E - V \over {T}} {d
x^i_{(m)} \over d \lambda}\sqrt{{T} \over E - V}d^i_{(m)}V = - {d
V \over d \lambda}.
\end{equation}
But this is not the end since $\lambda$ is still free. If we
choose it such that
\begin{equation}
 {{T} \over E - V} = 1 \Rightarrow {T} = E - V \label{2.6a}
\end{equation}
then Eqs.(\ref{2.3a}) and (\ref{2.5a}) become
\begin{equation}
p^i_{(m)} = m_{(m)}{d x^i_{(m)} \over d \lambda},\hspace{1.0cm} {d
p^i_{(m)} \over d\lambda} = -d^i_{(m)}V,
 \label{2.7a}
\end{equation}
and we recover Newton's second law with respect to this special
$\lambda$, which has the same properties as Newton's absolute
time. However, Eq.(\ref{2.6a}), which is usually taken to express
energy conservation, becomes, in the absence of an external time,
the definition of `Newtonian' time. Indeed, this emergent time is
the astronomers' operational ephemeris time \cite{bb}.

We now consider the `bad' (suffix B) Jacobi action with square
root and summation swapped:
\begin{equation}
    I_\textrm{B} =     2\int\sqrt{E -
V}\sum_{(m) = 1}^N\sqrt{{T}_{(m)}} d\lambda,
\end{equation}
where $E$, $V$ and $\lambda$ have their previous meanings, and
$$  {T}_{(m)} =  {m_{(m)} \over
2}{dx_{(m)}^i\over d\lambda}{dx_{(m)i}\over d\lambda}$$ is the
$m^{\textrm{th}}$ particle's kinetic energy.

We will generalise this slightly by bringing the `$E - V$' inside
the summation sign. Then each particle can have its own energy and
potential, and we have \begin{equation}
    I'_\textrm{B} = 2\int\sum_{(m) = 1}^N\sqrt{E_{(m)} -
V_{(m)}}\sqrt{{T}_{(m)}} d\lambda, \label{2.8aa}
\end{equation}

In Jacobi's action, the `distance' in configuration space is an
$N$-dimensional Pythagorean sum, i.e., the square root of a sum of
squares (with mass-weighted `legs'). In the new action, we add the
$ds$'s along a line. In the `good' $ds$, the change due to a
change in one leg depends also on what all the other particles are
doing. This linkage is not present in the `bad' $ds$.

The new canonical momenta are
\begin{equation}
    p^{i}_{(m)} = {\delta {\cal L} \over \delta\left({d x_{(m)i} \over
d \lambda}\right)} = m_{(m)}\sqrt{E_{(m)} - V_{(m)} \over
{T}_{(m)}} {d x^i_{(m)} \over d \lambda}.\label{2.9aa}
\end{equation}

Now to the point. Instead of just one in the `good' case, there
are now $N$ independent identities, one for each particle.
They arise just due to the placing of the square root. For this one
core mathematical relationship, we shall use various names:
reparametrization or square-root identity; Hamiltonian, quadratic
or square-root constraint. The new identities are
\begin{equation}
{p^{i}_{(m)}p_{(m)i} \over 2m_{(m)}} = {E_{(m)} - V_{(m)} \over
{T}_{(m)}} \times {T}_{(m)} = E_{(m)} - V_{(m)} . \label{2.4aa}
\end{equation}
The Euler--Lagrange equations
\begin{equation}
{d  p^{i}_{(n)} \over d \lambda} = {\delta {\cal L} \over \delta
x_{(m)i}} = -\sum_{(m) = 1}^N{\sqrt{T}_{(m)} d^i_{(n)}V_{(m)}\over
\sqrt{E_{(m)} - V_{(m)}}}
 \label{2.5aa}
\end{equation}
do not now in general preserve the constraints (\ref{2.4aa}).
There are extra secondary constraints, one for each particle:
\begin{eqnarray}
&{d \over d\lambda}\left[{p^{i}_{(m)}p_{(m)i} \over 2m_{(m)}} -
E_{(m)} + V_{(m)}\right] = {p_{(m)i}\over m_{(m)}}{d  p^{i} \over
d \lambda} + {d V_{(m)} \over d \lambda} =  -{p_{(m)i}\over
m_{(m)}} \sum_{(n) = 1}^N{\sqrt{T}_{(n)} d^i_{(m)}V_{(n)}\over
\sqrt{E_{(n)} - V_{(n)}}} + \sum_{(n) = 1}^N
 d^i_{(n)}V_{(m)}{d x_{(n)i} \over d \lambda}\nonumber\\
&= -{p_{(m)i}\over m_{(m)}} \sum_{(n) = 1}^N{\sqrt{T}_{(n)}
d^i_{(m)}V_{(n)}\over \sqrt{E_{(n)} - V_{(n)}}} + \sum_{(n) = 1}^N
 d^i_{(n)} V_{(m)}{ p_{i(n)} \over m_{(n)}}\sqrt{T_{(n)} \over E_{(n)} -
V_{(n)}}\label{2.6aa}
\end{eqnarray}
Here, we inverted the expression (\ref{2.9aa}) for the momenta to
find $d x_{(m)i}/d \lambda$ and substituted to obtain the last
line. This does not vanish identically and thus is a new secondary
constraint. We cannot have such proliferation, so are there any
cases for which the expressions (\ref{2.6aa}) vanish?

The easiest way to understand their implications is to consider
special initial data. Suppose particle (1) is moving and all the
others are instantaneously at rest. Take Eq.(\ref{2.6aa}) for $(m)
= (2)$.  If it vanishes
\begin{equation}
0 = d^i_{(1)}V_{(2)}{   p^{i}_{(1)} \over m_{(1)}}\sqrt{T_{(1)}
\over E_{(1)} - V_{(1)}}. \label{2.6aaa} \end{equation}
This requires $d^i_{(1)}V_{(2)} \equiv 0$. Taking different $(m)$ and
$(n)$, we find $d^i_{(m)}V_{(n)} \equiv 0 \forall n \ne m$, i.e.,
$V_{(n)}$ can only depend on $x^i_{(n)}$ and not on the other
particles' coordinates. Then from Eq.(\ref{2.6aa}) we find that
all the secondary constraints are identically satisfied, with no
restriction on $V_{(n)}$. The particles can each move in their own
external potential but cannot interact among themselves.

Physically, the system starts off with one global
reparametrisation invariance. However, each particle wants to be
independently reparametrisable. To see this, we seek an equation
to propagate the kinetic energy $T_{(m)}$ of any one particle.
None exists. The direction of each particle is fixed, but its
speed is arbitrary. To see this another way, note
[Eq.(\ref{2.9aa})] that the momenta are homogeneous of degree zero
in the velocities. They depend only on the direction of motion of
each particle and not at all on its speed. The Euler--Lagrange
equations (\ref{2.5aa}) govern the momenta, which are directions,
and thus say nothing about speeds. This is why the system makes
sense only when the particles do not interact. We may also note
that the local square-root action is the sum of $N$ independent
single-particle Jacobi actions. The particles cannot interact; but
if they do not interact each can be independently reparametrised.

The `bad' Hamiltonian is illuminating. It is just the sum of the
constraints, each with a Lagrange multiplier:
\begin{equation}
H_\textrm{B} = \sum_{(m) = 1}^N N_{(m)}\left({p^{i}_{(m)}p_{(m)i}
\over 2m_{(m)}} - E_{(m)} + V_{(m)} \right).\label{H_b}
\end{equation}
Variation w.r.t. the multipliers $N_{(m)}$ gives the constraints.
Hamilton's equations are
\begin{equation}
{d p^{i}_{(m)} \over dt} = -N_{(m)}d^i_{(m)}V_{(m)}, \hskip 1cm {d
x^{i}_{(m)} \over dt} = N_{(m)}\left({p^{i}_{(m)} \over
2m_{(m)}}\right).
\end{equation}

These are exactly the equations of motion (\ref{2.9aa}) and
(\ref{2.5aa}) from $I_\textrm{B}$ with the identification $N_{(m)}
= \sqrt{T_{(m)}/E_{(m)} - V_{(m)}}$. Once again the kinetic energy
is seen to be undetermined.

The `bad' action (\ref{2.8aa}) evidently has interesting and
important properties: The system is over-constrained and the
constraints do not, in general, propagate; the `defect' acts as a
filter, picking out the `simplest' action; the special action has
a `hidden' symmetry (not explicitly exhibited in the action); the
`bad' Hamiltonian is just a sum of the constraints.
Alternatively, one could start with a Hamiltonian such as
Eq.(\ref{H_b}) which is a sum of constraints, each of which is
quadratic in the (undifferentiated) momenta. Now all one has to
do is repeat the BSW reduction procedure. First one constructs a
Lagrangian by means of a Legandre transformation. This will
contain the Lagrange multipliers. These are eliminated and one
ends up back with Eq.(\ref{2.8aa}). All these properties carry over
to the extension of the `bad' Jacobi action (\ref{2.8aa}) to a field
theory, in particular to gravity.

\section{Reparametrisation-invariant actions and gravity}

As we already said at the start, 40 years ago Dirac \cite{dirac}
and Arnowitt, Deser and Misner (ADM) \cite{adm} showed that GR
could be treated as a dynamical system on the configuration space
of Riemannian 3-geometries with fixed topology (called
{\it{superspace}} by Wheeler\cite{jaw}). We work in this paper
with compact manifolds without boundary; the extension to more
general topologies (e.g., asymptotically flat) is straightforward.

We now begin the programme outlined in the introduction. We shall
consider a class of Jacobi-type actions on superspace. We will
show that the requirement that the equations of motion be strongly
local (in a sense to be defined) and self consistent drastically
restricts the actions.  In fact, standard GR with locally
Lorentz-covariant matter fields that interact through the gauge
principle emerges naturally.  We begin with matter-free superspace
and obtain pure geometrodynamics.

For definiteness, we take the compact manifold to be $S^3$.  We
will actually work in {\it Riem}, the space of suitably smooth
Riemannian metrics $g_{ij}(x), x \in S^3$.  The equivalence
classes \{$g_{ij}$\} of the metrics $g_{ij}$ under suitably smooth
3-diffeomorphisms on {\it Riem} are the points in superspace.
Consider a smooth family of smooth diffeomorphisms $x'^i(x^j,
\theta)$, where the label $\theta$ distinguishes the members of the
family and the diffeomorphism goes to the identity map as $\theta
\rightarrow 0$. The general infinitesimal 3-diffeomorphism of
$g_{ij}(x)$ is generated by the Killing form by
\begin{equation}
    {\partial g_{ij}'\over \partial \theta}|_{\theta = 0} =
{\cal L}_{\xi}g_{ij} = (K\xi)_{ij}
 = \nabla_i\xi_j + \nabla_j\xi_i,\label{kf}
\end{equation}
where $\xi^i(x) = \partial x'^i/ \partial\theta$ evaluated at
$\theta = 0$, and ${\cal L}$ indicates the Lie derivative. The
Killing form, with its multi-index covariant derivatives
$\nabla_{j}$, is the most intricate Lie derivative we shall
encounter; propagation of its intricacy creates gravity and
imposes deep universal structure on matter coupled to gravity. The
Killing form drives it all through the square-root filter.

We seek a measure on superspace like the `bad' Jacobi action
(\ref{2.8aa}). Consider two metrics on, say, $S^3$, which we call
$g^{(1)}$ and $g^{(2)}$. We take coordinates with the same range,
so that we can identify points on the two manifolds with the same
coordinate values. We can compute, at each point, the difference
between the metrics as $dg_{ab} = g^{(2)}_{ab} - g^{(1)}_{ab}$,
where we take $dg$ to be small but finite. We want a difference
linearly proportional to $dg$, something like $\sqrt{(dg)^2}$.
There are two natural squares of a symmetric two-index tensor. One
is the sum of the squares of the individual elements, while the
other is the square of the trace. We include both with an as yet
undetermined coefficient, and, maintaining maximum generality with
strong locality, we consider the object
\begin{equation}
    \tau = G^{abcd}[dg_{ab}][dg_{cd}],
\end{equation}
\begin{equation}
    G^{abcd} =g^{ac}g^{bd} - A g^{ab}g^{cd} \label{G}
\end{equation}
where $A$ is the undetermined constant coefficient. Clearly,
$\sqrt{\tau}$ is a natural measure of the difference of $g^{(1)}$
and $g^{(2)}$ at one coordinate point. To find the total
difference between $g^{(1)}$ and $g^{(2)}$, we need to integrate
over $S^3$. To obtain a more general measure, we can modulate
$\sqrt{\tau}$ by some spatial function of the metric, which we
call $\sqrt{P}$, to give
\begin{equation}
\Delta(g^{(1)}, g^{(2)}) = \int\sqrt{g}\sqrt{P}\sqrt{\tau}d^3x.
\label{D}
\end{equation}
Let us assume that all the unlabelled $g$'s are $g^{(1)}$'s and
integrate over the first manifold. The second manifold is an
equally good base that gives an answer whose slight difference is
immaterial in the limit. It is clear that ({\ref{D}}) is the
natural generalization to a field theory of the `bad' Jacobi
action (\ref{2.8aa}). We have merely replaced summation over
particles by integration over a field.

This is a distance on {\it Riem}. To get a distance on superspace,
we make different coordinate transformations on each of the two
manifolds, getting $g'^{(1)}$ and $g'^{(2)}$. We recalculate
$\Delta(g'^{(1)}, g'^{(2)})$ and, in the key step, find the
minimum of $\Delta$ over all possible transformation pairs. The
minimum, provided it exists, will be a function on superspace. In
fact, we only need to make transformations on the second manifold,
because identical transformations on the two manifolds leaves
$\Delta$ unchanged.

To transform this finite-distance function $\Delta$ into an
infinitesmal, consider a curve of metrics with label $\lambda$
that interpolates $g^{(1)}$ and $g^{(2)}$. Evaluate $\partial
\Delta /
\partial \lambda$ at any $\lambda$ and get
$$ {\partial \Delta \over \partial \lambda} =
\int\sqrt{g}\sqrt{P}\sqrt{{T}}d^3x,$$ with \begin{equation}
    {T} = G^{abcd}\left[{\partial g_{ab} \over \partial
\lambda}\right]\left[{
\partial g_{cd}
\over \partial \lambda}\right],
\end{equation}
with $G^{abcd}$ given by (\ref{G}).

To take into account the $\lambda$-dependent coordinate
transformations, we add a Killing form to
$ \partial g_{ab} /
\partial \lambda$, $ \partial g_{ab} /
\partial \lambda\rightarrow \partial g_{ab} /
\partial \lambda - (K\xi)_{ab}$, where $\xi$ is {\it a priori} an
arbitrary 3-vector field. The `distance' along the curve between
$g^{(1)}$ and $g^{(2)}$ is now
\begin{equation}
    \Delta(g^{(1)}, g^{(2)}) = {\rm extremum~with~respect~to} ~\xi
~{\rm of} \int d\lambda\int d^3x\sqrt{g}\sqrt{P}\sqrt{T}.
\label{1.6a}
\end{equation}

The extremalization in (\ref{1.6a}) w.r.t. $\xi$, which we call
{\it{best matching}} \cite{bb}, makes $\Delta$ into a measure on
superspace and not just on {\it Riem}. We use the word
{\it{measure}}, as opposed to metric, because we are actually about
to create a degenerate (albeit highly interesting) structure. The
ideal, motivated in \cite{bb} by an idea that goes back to
$\textrm{Poincar}\acute \textrm{e}$, would be to have a true
metric on superspace with geodesics as dynamical curves. However,
because we choose the `bad' action, we will find that although a
best-matching action is in principle defined on all curves in
superspace there are many curves joining two given points in
superspace that have the same action. This degeneracy would not
have occurred had we opted for the `good' action. The consequence
of our `bad' choice coupled with the square-root filter is to
acquire an extra symmetry, which, unlike the 3-diffeomorphism
symmetry on the configuration space, acts on the phase space.

It is worth relating this to the number of physical degrees of
freedom of the gravitational field, which is universally agreed to
be two if one rules out tensor--scalar gravity of Brans--Dicke
type. Now, since the $3\times 3$ symmetric tensor $g_{ij}$ has six
independent components of which three correspond to coordinate
freedom, a true geodesic theory on superspace must have three
physical degrees of freedom per space point. Our `bad' choice
introduces an extra scalar symmetry and takes us down to two
physical degrees of freedom. Our readers might begin to suspect
that we are merely taking them on a roundabout route to standard
GR.

To counter this worry, we first note that even if it is indirect
the way is new. We get to the goal with much reduced kinematics --
there is no time and no presupposed Minkowskian spacetime
structure in the small. More significantly, we get new insights
into the origin of the universal light cone and gauge theory. We
should also like to draw attention to our earlier paper
\cite{bom}, in which we apply best matching and a local square
root on conformal superspace (CS). Since the 3-vector symmetry of
the 3-diffeomorphisms is augmented on CS by the scalar symmetry of
3-conformal transformations, the six degrees of freedom on {\it
Riem} are directly reduced to 6 - 3 - 1 = 2 on CS. Remarkably, the
use of the local square root in this case still acts as a filter of
theories but does not reduce the physical degrees of freedom any
further. There is no extra phase-space symmetry associated with the
theory on CS. Since the coupling to matter on CS has still to be
worked out in detail, we shall not explore this route further in the
present paper. But it does hint at interesting possibilities (see also
\cite{barb}).

Returning to the task in hand, the variational principle on
superspace, the final step is to seek the curve between $g^{(1)}$
and $g^{(2)}$ that minimizes the best-matching `distance' between
the two geometries . This is achieved by varying w.r.t. the metric
$g_{ij}$.

Note that ${T}$ as written is the most general ultralocal (i.e.,
with the supermetric $G^{abcd}$ dependent only on $g_{ij}$ and not
on its spatial derivatives) quadratic form in $dg_{ij}$ up to an
overall constant (which cannot affect the resulting geodesic
curves).  As we noted, the first term is the sum of squares and
the second the square of the trace.  The relative contributions of
these two scalar terms has yet to be determined, hence the
coefficient $A$. For $A = +1$, $G^{abcd}$ is the DeWitt
supermetric.

The above analysis shows that $\xi$ in the action is the rate of
change of a three-dimensional coordinate transformation, i.e., it
is a `velocity'. Moreover, the action does not depend on the
associated `position' (three-dimensional coordinate
transformation). Thus, it appears that $\xi$ should be treated as
a `cyclic' or `ignorable' coordinate \cite{lanczos}, rather than
as a Lagrange multiplier (a `position' without a `velocity'). In
fact, it will be shown in \cite{barb} that $\xi$, like all such
variables used to implement best matching, is neither a multiplier
nor a cyclic coordinate but a {\it{sui generis}} variable for
which the variation at the end points is not fixed. The same
analysis will show that in most cases, including the one
considered here, the treatment as a multiplier is valid, though
there are cases in which it is not. In this paper, we shall
therefore regard $\xi$ as a multiplier, which matches the standard
treatment in gauge theory.

We conclude this section by underlining the presence of the local
square root in (\ref{1.6a}), i.e., the square roots are taken
before the integration over space. This is the analogue of the
`bad' Jacobi action. The analogue of the `good' action with
`global' square roots is
\begin{equation}
I = \sqrt{\int d^3x \sqrt{g}P}\sqrt{\int d^3x
\sqrt{g}{T}}.\label{glob}
\end{equation}

Such an action yields geometrodynamics with three freedoms per
point. Moreover, the global square roots do not `filter'. The one
basic form (\ref{glob}) allows many instantiations. In contrast,
mere consistency applied to candidates for $P$ and $T$ in
(\ref{1.6a}) and its obvious generalization to scalar and 3-vector
fields on {\it Riem} leads directly to matter-free gravity (Sec. V), to
matter fields minimally coupled to gravity and obeying its light
cone (Secs. VI and VII) and to the gauge principle and
massless electrodynamics (Sec. VII). These are all latent
in (\ref{1.6a}). We must now make them appear.

\section{The BSW Action and Its Hamiltonian Form}

The Baierlein--Sharp--Wheeler (BSW) action \cite{bsw} has the form
\begin{equation}
A_{\textrm{BSW}} = \int d\lambda \int \sqrt{g} \sqrt{R} \sqrt{{T}}
d^3x, \label{2.1}
\end{equation}
where the `kinetic energy' $T$ is
\begin{equation}
T = (g^{ac}g^{bd} - g^{ab}g^{cd}) \left[{\partial g_{ab} \over
\partial \lambda} - (K\xi)_{ab}\right]\left[{\partial g_{cd} \over
\partial\lambda} - (K\xi)_{cd}\right],
\end{equation}
the monotonic $\lambda$ labels the 3-metrics $g_{ij}(x, \lambda)$
on a curve in {\it Riem}, $R$ is the scalar curvature of $g_{ij}$, and
the other symbols have been explained.

Although it was not found in this way, the BSW action can clearly
be derived from the `distance' (\ref{1.6a}) with $A = +1$ and $R$
as the `potential' $P$.  We first show that $A_{\textrm{BSW}}$
leads to the Dirac--ADM Hamiltonian. The Lagrangian density is
${\cal L} = \sqrt{g}\sqrt{R} \sqrt{{T}}$, and the canonical
momenta conjugate to $g_{ij}$ are
\begin{equation}
    p^{ij} = {\delta {\cal L} \over \delta\left({\partial g_{ij} \over
\partial \lambda}\right)} = \sqrt{gR \over {T}}(g^{ic}g^{jd} -
g^{ij}g^{cd}) \left[{\partial g_{cd} \over \partial\lambda} -
(K\xi)_{cd}\right]. \label{2.2}
\end{equation}

We immediately observe their homogeneity of degree zero in the
`corrected velocities' $\partial g_{ab}/\partial\lambda -
(K\xi)_{ab}$, which occur linearly in the numerator and
quadratically under the square root in the denominator $T$.  They
define a [`local'] {\it{direction}} in superspace, as opposed to
the direction and speed of ordinary momenta. The BSW action is
therefore timeless, determining only paths without speed in
superspace. Just as squared direction cosines sum to 1, the BSW
canonical momenta satisfy the {\it{square-root identity}}
\begin{equation}
    gH = -p^{ij}p_{ij} + {1 \over 2}p^2  + gR = 0, \label{2.3}
\end{equation}
with the trace $p = g_{ab}p^{ab}$. This identity is obtained by
substitution of (\ref{2.2}) into (\ref{2.3}) and in generalized
Hamiltonian dynamics is a primary constraint \cite{dirac}. Another
primary constraint is the identical vanishing of the momentum
conjugate to $\xi^i$ that arises from our taking it to be a
multiplier. Then variation of (\ref{2.1}) w.r.t $\xi^i$ as a
multiplier gives
\begin{equation}
    2\left(\sqrt{gR \over {T}}(g^{ic}g^{jd} - g^{ij}g^{cd})
\left[{\partial g_{cd} \over \partial\lambda} - (K
\xi)_{cd}\right]\right)_{;i} = 0. \label{2.4}
\end{equation}
This, on using Eq.(\ref{2.2}), gives us the secondary constraint
\begin{equation}
 \sqrt{g}H^j = p^{ij}_{~~;i} = 0, \label{2.5}
\end{equation}
which consists of three conditions per space point. We also have
the as yet undetermined vector $\xi$, with three components per
point. We wish to interpret (\ref{2.4}) as an equation for $\xi$.
The solution of this equation is the well-known {\it{thin-sandwich
problem}} \cite{ts}. Suppose we can find $\xi$ as a function of
$g_{ij}$ and $\partial g_{ij}/\partial \lambda$. We can then
substitute this back into (\ref{2.1}) to get a measure on
superspace that is determined solely by the 3-metric. The
auxiliary multiplier $\xi$ will have been eliminated, and we
should have a well-posed initial-value problem for $g_{ij}$ and
its $\lambda$ derivative.

This is what the logic of the 3-space approach suggests. In fact,
the thin-sandwich problem is burdened with difficulties \cite{ts},
and one should see our Lagrangian as heuristic rather than
practical. Fortunately, it leads unambiguously to a Hamiltonian
form -- the Dirac--ADM fully constrained Hamiltonian -- that is
much more tractable and in which, as York showed \cite{york}, one
can genuinely obtain a well-posed initial-value problem. Our motto
is therefore: ``Conceptualize in the configuration space,
calculate in the [constrained] phase space."

The entire evolution dynamics is in the Euler--Lagrange equations
for $g_{ij}$:
\begin{equation}
{\partial   p^{ij} \over \partial \lambda} = {\delta {\cal L}
\over \delta g_{ij}} = -\sqrt{g{T}\over 4R}\left(R^{ij} -
g^{ij}R\right) - \sqrt{{T} \over gR}\left(p^{im}p_m^{~j} - {1
\over 2}p~p^{ij}\right) +\left(\sqrt{g{T} \over 4R}^{;ij} -
g^{ij}\nabla^2\sqrt{gT \over 4R}\right) +{\cal L}_{\xi^i}p^{ij},
\label{2.6}
\end{equation}
where ${\cal L}_{\xi^i}$ stands for the Lie derivative along
$\xi^i$.

The square-root identity forces the standard Hamiltonian to vanish
identically, as it does for all Lagrangians homogeneous of degree
one in the velocities.  Using Dirac's generalized Hamiltonian
dynamics \cite{dirac}, we consider
\begin{equation}
{\cal H} = \int \sqrt{g}(NH + N_iH^i)d^3x \label{2.7}
\end{equation}
where $N$ and $N_i$ are position-dependent multipliers, and $H$
and $H^i$ are the constraints (\ref{2.3}) and (\ref{2.5}),
respectively. Variation w.r.t. $N$ and $N_i$ imposes $H = 0$ and
$H^i = 0$.

The expression (\ref{2.7}) is exactly the Dirac--ADM Hamiltonian;
$H$ is the Hamiltonian constraint; $H^i$ is the momentum
constraint; $N$ is the lapse; $N^i$ is the shift. The standard
equations of motion that it yields are
\begin{eqnarray}
    {\partial g_{ab} \over \partial \lambda} &=& 2{N \over
\sqrt{g}}\left(p_{ab} - {1 \over 2}g_{ab}p\right) +N_{a;b} +
N_{b;a}\label{2.9}\\
{\partial p^{ij} \over \partial \lambda} &=& -\sqrt{g}N\left(R^{ij} - {1
\over 2}g^{ij}R\right) +{Ng^{ij} \over 2 \sqrt{g}}\left(p^{ab}p_{ab} -{1
\over 2}p^2\right) - {2N \over \sqrt{g}}\left(p^{im}p_m^{~j} - {1
\over 2}p~p^{ij}\right)\nonumber\\& & +\sqrt{g}\left(N^{;ij} -
g^{ij}\nabla^2N\right) +{\cal L}_{N^i}p^{ij}.\label{2.10}
\end{eqnarray}

Inverting (\ref{2.2}) to obtain $\partial g_{ij} /
\partial \lambda$ in terms of the momenta,
\begin{equation}
{\partial g_{ij} \over \partial \lambda} = \sqrt{T\over
gR}\left(p_{ij} - {1 \over 2}g_{ij}p\right) +\xi_{i;j}
+\xi_{j;i},\label{2.11}
\end{equation}
and comparing (\ref{2.6}) with (\ref{2.10}) and (\ref{2.11}) with
(\ref{2.9}), we get the identifications
\begin{equation}
N = \sqrt{T\over 4R}, ~~N^i = \xi^i. \label{2.12}
\end{equation}

In fact, even after these identifications, the dynamical equations
(\ref{2.6}) and (\ref{2.10}) differ by a multiple of the
Hamiltonian constraint. This has no conceptual significance
because the constraints vanish, but such differences can be
important in numerical work. Moreover, if in the ADM Hamiltonian
one chooses $N/\sqrt{g}$ instead of $N$ as the independent
variable, then, as emphasized by York \cite{jwy}, the extra term
vanishes, and one gets complete agreement. This shows that the BSW
and ADM solution curves are closely related.

However, they are not quite identical. The ADM lapse and shift are
entirely free functions, while the BSW logic calls for the
solution of (\ref{2.4}) for $\xi$. If this succeeds, $\sqrt
{T/4R}$ is determined. It seems that the BSW initial data
determine a unique direction in superspace as distinct from the
ADM `many-fingered time'. This apparent discrepancy will be
resolved in Sec. V.

\section{  Propagation of the BSW Constraints}

For consistency, the evolution equations of constrained theories
must propagate the constraints \cite{dirac}: initially zero, they
must remain so.  Thus, we must have $\partial H /\partial \lambda
= 0$ and $\partial H^i/\partial \lambda = 0$ by virtue of
(\ref{2.6}) and the definition (\ref{2.2}). Since constraint
propagation is crucial throughout this paper, we shall exhibit it
in some detail for BSW.  We begin with the momentum constraint
$H^i = 0$, differentiating it w.r.t. $\lambda$ and using the BSW
(not ADM) evolution equations (\ref{2.6})--(\ref{2.11}) to replace
the $\lambda$ derivatives of $g_{ij}$ and $p^{ij}$.  To simplify
(and make the ADM connection), we write $N$ in place of $\sqrt
{{T}/4R}$. After cancellations and rearrangements, we are left
with
\begin{eqnarray}
{\partial \over \partial \lambda}\left(p^{ij}_{~~;j}\right) =& {1
\over 2} N\left[\sqrt{g}R - {1 \over \sqrt{g}}\left( p^{ab}p_{ab}
- {1 \over 2}p^{2}\right)\right]^{;i}
 + N^{;i}\sqrt{g}\left[R - g^{-1}\left(p^{ab}p_{ab} - {1 \over
2}p^{2}\right)\right] \nonumber \\ &- {2N \over \sqrt{g}}\left(p^{im}
- {1 \over 2}g^{im} p\right)p^b_{~m;b} + {\cal
L}_{\xi}\left(p^{ij}_{~~;j}\right). \label{3.1} \end{eqnarray}

Thus, $\partial/\partial\lambda(p^{ij}_{~~;j})$ vanishes weakly
\cite{dirac}: if the constraint holds initially, it will
propagate. For the Hamiltonian constraint
\begin{equation}
{\partial \over \partial \lambda}\left[gR - \left( p^{ab}p_{ab} - {1 \over
2}p^{2}\right)\right] = 4N^{;a}p^b_{~a;b} + 2Np^{ab}_{~~;ab} + {\cal
L}_{\xi}\left[gR - \left( p^{ab}p_{ab} - {1 \over
2}p^{2}\right)\right],
\label{3.2}\end{equation}
 so this constraint too propagates. (In
actual fact, when expressed in Lagrangian, as opposed to
Hamiltonian, form, the expressions within the square parentheses
in (\ref{3.2}) vanish identically. This vanishing is precisely the
content of the square-root identities. However, both here and in
the equations below, we prefer to show these terms explicitly, if
for no other reason than that they must be present in the
Hamiltonian formulation, in which, as we already noted, the
Lagrangian identity becomes a true Hamiltonian constraint.)

In the light of the definition (\ref{2.2}), the evolution
equations (\ref{2.6}) appear to give six equations for the six
$\partial^2g_{ij}/\partial \lambda^2$. The fact that (\ref{2.2})
propagate the four constraints tells us that only two of
(\ref{2.2}) are true evolution equations. Therefore, the system is
highly under-determined. In particular, we expect that $\partial
(p^{ij}_{~~;j})/
\partial \lambda = 0$ should be used to solve for $\partial \xi /
\partial \lambda$. Since the constraint is preserved, we end up with
no restriction on $\partial \xi /
\partial \lambda$. A similar situation is true for $\partial N /
\partial \lambda$.

The upshot is that $N$ and $\xi$ are fixed on the initial slice by
the initial data, but their evolution is free. Hence the freedom
in the ADM lapse and shift is effectively shared by the BSW
action, and both generate Einsteinian gravity.

In the local-square-root particle model, discussed in Section II, in the
special case where the constraints are propogated the action reduces to a
sum of single particle Jacobi actions. Therefore the action is locally
reparametrisation invariant even though it is expressed in terms of a
glabal parameter. This is {\it not} true in the BSW action. The spatial
derivatives in the scalar curvature are evaluated at a fixed `time'. If we
make a `local' parameter change we will change the surfaces of constant
time and thus change the derivatives in a complicated fashion. Therefore
the BSW action remains only globally reparametrisation invariant.

\section{Uniqueness of BSW}

We have seen that constraint propagation is important. Many have
sought conditions under which GR can be derived. Two main
strategies have been followed. The older classical arguments,
reviewed by Hojman, Kucha\v r, and Teitelboim (HKT) \cite{hkt1},
relied on four-dimensional general covariance coupled with
simplicity restrictions in a Lagrangian framework. These
essentially select the Hilbert action uniquely (up to an arbitrary
cosmological constant).  More recently, Teitelboim \cite{t}
started from a Hamiltonian viewpoint and deduced matter-free GR by
postulating: 1) that the Hamiltonian should have the local form
(\ref{2.7}); 2) that $H$ and $H^i$ should depend only on the
3-metric $g_{ij}$ and its conjugate momentum $p^{ij}$; and 3) that
the resulting dynamics should satisfy an embeddability criterion
proposed by Wheeler: ``If one did not know the
Einstein--Hamilton--Jacobi equation, how might one hope to derive
it straight off from plausible first principles, without ever
going through the formulations of the Einstein field equations
themselves?  The central starting point in the proposed derivation
would necessarily seem to be `embeddability' [in a
four-dimensional pseudo-Riemannian spacetime]."

    As Teitelboim noted in his PhD thesis \cite{tt}, this is an extremely
restrictive condition.  Developing an approach of Dirac
\cite{dirac}, he showed that embeddability imposes a strict
requirement on the Poisson-bracket relations between $H$ and
$H^i$. They must satisfy the so-called Dirac algebra.  In
\cite{hkt1}, HKT then sought theories in which the manner in which
the constraints close ensures embeddability and showed (again with
certain simplicity requirements) that GR is the unique theory that
does so.

As our first new result, we show that embeddability is a much
stronger condition than one needs.  The constraint algebra need
not close in a specific way.  It is merely necessary that it
close. As we shall see, this opens up an entirely new derivation
of relativity -- both the special and the general theory -- in
which no {\it a priori} assumption of geometrodynamic evolution of
spacelike hypersurfaces in a four-dimensional pseudo-Riemannian
spacetime is made.  We can derive relativity without relativity
merely by postulating an action based on a metric `distance' of
the form (\ref{1.6a}) and requiring that its constraints
propagate.

Best matching, which ensures 3-diffeomorphism invariance,
automatically leads to a momentum constraint of the form
(\ref{2.5}). The local square root leads to a local square-root
identity like (\ref{2.3}), which becomes a quadratic Hamiltonian
constraint like (\ref{2.4}). We are led naturally to a local
Hamiltonian of the form (\ref{2.7}). Both constraints strongly
restrict the possible Lagrangians (or Hamiltonians) through the
condition of constraint propagation.

We make no attempt at an exhaustive analysis and employ a
pedestrian technique.  We suspect our various individual results
could be obtained more elegantly in a unified manner but think it
premature to seek it at this stage, since there are several
extensions of the method, which we shall mention at the end of the
paper, that should first be explored. In the meanwhile, our
individual results show the potential of the 3-space approach.

We start with the simplest modification of the BSW Lagrangian:
changing the coefficient $A$ in the supermetric from the DeWitt
value $A = 1$. The inverse to the supermetric is $g_{ae}g_{bf} -
{A \over 3A - 1} g_{ab}g_{ef}$ because
\begin{equation}
\left[g_{ae}g_{bf} - {A \over 3A - 1}
g_{ab}g_{ef}\right]\left[g^{ac}g^{bd} - Ag^{ab}g^{cd}\right] =
\delta^c_e \delta^d_f.
\end{equation}
We define
\begin{equation}
B = {2A \over 3A-1}
\end{equation}
because when $A = 1$ we also have $B = 1$.

Hence we start with a modified BSW action
\begin{equation}
A'_{\textrm{BSW}} = \int d\lambda \int \sqrt{g} \sqrt{R} \sqrt{T}
d^3x
\end{equation}
with
\begin{equation}
T = (g^{ac}g^{bd} - Ag^{ab}g^{cd}) \left[{\partial g_{ab} \over
\partial \lambda} - (K\xi)_{ab}\right]\left[{\partial g_{cd} \over
\partial\lambda} - (K\xi)_{cd}\right].
\end{equation}
The canonical momenta conjugate to $g_{ij}$ are
\begin{equation}
    p^{ij} = {\delta {\cal L} \over \delta\left({\partial g_{ij} \over
\partial \lambda}\right)} = \sqrt{gR \over T}(g^{ic}g^{jd} - Ag^{ij}g^{cd})
\left[{\partial g_{cd} \over \partial\lambda} -
(K\xi)_{cd}\right].
\end{equation}
This can be inverted to give\begin{equation} {\partial g_{ij}
\over \partial \lambda} = \sqrt{T \over gR}\left(p_{ij} - {B \over
2}g_{ij}p\right) +\xi_{i;j} +\xi_{j;i}.\label{4.6}
\end{equation}
The square-root identity becomes
\begin{equation}
    gH = -p^{ij}p_{ij} + {B \over 2}p^{2}  + gR = 0, \label{4.7}
\end{equation}
while the linear constraint that arises from varying w.r.t. $\xi$
is unchanged:
\begin{equation}
 \sqrt{g}H^j = p^{ij}_{~~;i} = 0. \label{4.8}
\end{equation}
The evolution is again in the Euler--Lagrange equations for
$g_{ij}$:
\begin{equation}
{\partial   p^{ij} \over \partial \lambda} = {\delta {\cal L}
\over \delta g_{ij}} = -\sqrt{gT \over 4R}\left(R^{ij} -
g^{ij}R\right) - \sqrt{T \over gR}\left(p^{im}p_m^{~j} - {B \over
2}pp^{ij}\right) +\left(\sqrt{gT \over 4R}^{;ij} -
g^{ij}\nabla^2\sqrt{gT \over 4R}\right) +{\cal L}_{\xi^i}p^{ij}.
\label{4.9}
\end{equation}
We use equations (\ref{4.6}) and (\ref{4.9}) to evolve the
constraints, obtaining
\begin{eqnarray}
{\partial \over \partial \lambda}\left(p^{ij}_{~~;j}\right) =& {1
\over 2} N\left[\sqrt{g}R - {1 \over \sqrt{g}}\left( p^{ab}p_{ab}
- {B \over 2}p^{2}\right)\right]^{;i}
 + N^{;i}\sqrt{g}\left[R - g^{-1}\left(p^{ab}p_{ab} - {B \over
2}p^{2}\right)\right] \nonumber \\ &- {2N \over
\sqrt{g}}\left(p^{im} - {B \over 2}g^{im}p\right)p^b_{~m;b} +
{\cal L}_{N^i}\left((p^{ij}_{~~;j}\right). \label{4.10}
\end{eqnarray} The $\lambda$ derivative of the momentum
constraint, being proportional to itself, vanishes weakly, and so
the constraint propagates. However, for the Hamiltonian constraint
\begin{eqnarray}
{\partial \over \partial \lambda}\left[\sqrt{g}R - {1 \over
\sqrt{g}}\left( p^{ab}p_{ab} - {B \over 2}p^{2}\right)\right] =&
4N^{;a}p^b_{~a;b} + 2Np^{ab}_{~~;ab}
 + \left({3B - 2\over 2}\right) Np
\left[\sqrt{g}R - {1 \over \sqrt{g}}\left( p^{ab}p_{ab} - {B \over
2}p^{2}\right)\right]\nonumber \\
&+ \left(2B - 2\right)N\nabla^{2}p + \left(4B - 4\right)
N^{;i}p_{;i}\nonumber \\   &+ {\cal L}_{N^i}\left[\sqrt{g}R - {1
\over \sqrt{g}}\left( p^{ab}p_{ab} - {B \over
2}p^{2}\right)\right]. \label{4.11}\end{eqnarray} The right hand
side of (\ref{4.11}) does not vanish weakly. It is clear that the
trace $p$ must satisfy the secondary constraint \begin{equation} p
= \textrm{constant}. \label{CS}\end{equation} This is the
well-known constant-mean-curvature (CMC) gauge condition
\cite{york}, and it severely restricts the initial data. It also forms
a second-class constraint \cite{dirac} with the Hamiltonian, with
which it does not commute. When we evolve this constraint, we get the
standard CMC slicing condition:
\begin{equation}
\nabla^2N - RN = C,
\end{equation}
where $C$ is a spatial constant, essentially half $\partial
p/\partial \lambda$. This is yet another restriction on the
initial data, and these results show that we cannot connect
arbitrary 3-geometries, i.e., points on superspace, by curves that
extremalize the action based on the measure (\ref{1.6a}) unless $A
= 1$.

Thus, a consistent BSW-type action with $A\neq 1$ does not exist
on superspace. We note, however, that the secondary constraints
just obtained, particularly the special case $p=0$ of (\ref{CS}),
arise naturally if gravity is treated on conformal superspace (CS)
{\cite{york},\cite{bom}}. We have already mentioned our belief
that gravity will only be properly understood when treated as a
true geodesic theory on CS. We proceeded directly in that
direction in \cite{bom}, but we have since realized that the
difficulty presented by the condition (\ref{CS}) can be overcome
in two different ways: either on superspace with $A = 1$ or on CS
with (\ref{CS}) imposed \cite{bom}. Characteristically, both
`escape routes' are indicated by our method. We also mention that
our $A \neq 1$ result was already obtained by Giulini some years
ago \cite{g}.

We have seen that the sole apparent freedom (the value of $A$) in
the kinetic term $T$ is illusory, and that best matching and
consistency fix it uniquely. We now apply the same technique to
the `potential' $P$. One modification works. We first show that $P
= \Lambda + D R$ gives a consistent theory with $\Lambda$ as the
cosmological constant and $D$ is another constant.

Keeping the uniquely determined DeWitt $T$, we now modify the BSW
action to
\begin{equation}
A'_{\textrm{BSW}} = \int d\lambda \int \sqrt{g} \sqrt{DR + \Lambda}
\sqrt{T} d^3x.
\end{equation}
The momenta conjugate to $g_{ij}$ are
\begin{equation}
    p^{ij} = {\delta {\cal L} \over \delta\left({\partial g_{ij} \over
\partial \lambda}\right)} = \sqrt{g(DR + \Lambda) \over
T}(g^{ic}g^{jd} - g^{ij}g^{cd}) \left[{\partial g_{cd} \over
\partial\lambda} - (K\xi)_{cd}\right].
\end{equation}
This can be inverted to give\begin{equation} {\partial g_{ij}
\over \partial \lambda} = \sqrt{T \over g(DR +
\Lambda)}\left(p_{ij}  - {1 \over 2}g_{ij}p\right) +\xi_{i;j}
+\xi_{j;i}.\label{4.16}
\end{equation}
The square-root identity becomes
\begin{equation}
    gH = -p^{ij}p_{ij} + {1 \over 2}p^{2}  + g(DR + \Lambda)= 0,
\label{4.17}
\end{equation}
and we get the standard momentum constraint
\begin{equation}
 \sqrt{g}H^j = p^{ij}_{~~;i} = 0. \label{4.18}
\end{equation}
The Euler--Lagrange equations are
\begin{eqnarray}
{\partial   p^{ij} \over \partial \lambda} = {\delta {\cal L}
\over \delta g_{ij}} =& -\sqrt{gT \over 4(DR +
\Lambda)}\left(DR^{ij} - g^{ij}DR - g^{ij}\Lambda\right) - \sqrt{T
\over g(DR + \Lambda)}\left(p^{im}p_m^{~j} - {1 \over
2}pp^{ij}\right) \nonumber \\ &+\left(D\sqrt{gT \over 4(DR +
\Lambda)}^{;ij} - g^{ij}D\nabla^2\sqrt{gT \over 4(DR +
\Lambda)}\right) +{\cal L}_{\xi^i}p^{ij}. \label{4.19}
\end{eqnarray}
We use (\ref{4.16}) and (\ref{4.19}) to evolve the constraints,
noting that now $N = \sqrt{T/4(DR + \Lambda)}$. We get
\begin{eqnarray}
{\partial \over \partial \lambda}\left(p^{ij}_{~~;j}\right) =& {1
\over 2} N\left[\sqrt{g}(DR + \Lambda) - {1 \over \sqrt{g}}\left(
p^{ab}p_{ab} - {1 \over 2}p^{2}\right)\right]^{;i}
 + N^{;i}\sqrt{g}\left[DR + \Lambda - g^{-1}\left(p^{ab}p_{ab} - {1
\over 2}p^{2}\right)\right] \nonumber \\ &- {2N \over
\sqrt{g}}\left(p^{im} - {1 \over 2}g^{im}p\right)p^b_{~m;b} +
{\cal L}_{N^i}\left((p^{ij}_{~~;j}\right). \label{4.20}
\end{eqnarray} Thus, as expected, the momentum constraint
propagates. For the Hamiltonian constraint we have
\begin{eqnarray}
&{\partial \over \partial \lambda}\left[\sqrt{g}(DR + \Lambda)- {1
\over \sqrt{g}}\left( p^{ab}p_{ab} - {1 \over
2}p^{2}\right)\right] =\nonumber\\ & 4DN^{;a}p^b_{~a;b} +
2DNp^{ab}_{~~;ab}
 +{1 \over 2} Np
\left[\sqrt{g}(DR + \Lambda) - {1 \over \sqrt{g}}\left(
p^{ab}p_{ab} - {1
\over 2}p^{2}\right)\right]\nonumber \\
&+ {\cal L}_{N^i}\left[\sqrt{g}(DR + \Lambda) - {1 \over
\sqrt{g}}\left( p^{ab}p_{ab} - {1 \over 2}p^{2}\right)\right].
\label{4.21}\end{eqnarray} The right-hand side of (\ref{4.21})
also vanishes weakly.

As we mentioned in the introduction, we can by multiplying by an
overall scaling factor set $D = +1, -1,$ or 0. The first case is
standard general relativity with a cosmological constant, the second
case is euclidean gravity, again with a cosmological constant, and the
third is strong gravity.

We now try to modify $P$ more radically. Keeping to our pedestrian
approach, we consider two special but illuminating cases. The
first is $P = R^{\alpha}$ with $\alpha$ a constant. Then, with the
DeWitt $T$, we have
\begin{equation}
A'_{\textrm{BSW}} = \int d\lambda \int \sqrt{g} \sqrt{R^{\alpha}}
\sqrt{T} d^3x.
\end{equation}
The momenta conjugate to $g_{ij}$ are
\begin{equation}
    p^{ij} = {\delta {\cal L} \over \delta\left({\partial g_{ij} \over
\partial \lambda}\right)} = \sqrt{gR^{\alpha} \over T}(g^{ic}g^{jd} -
g^{ij}g^{cd}) \left[{\partial g_{cd} \over \partial\lambda} -
(K\xi)_{cd}\right].
\end{equation}
The inversion gives \begin{equation} {\partial g_{ij} \over
\partial \lambda} = \sqrt{T \over gR^{\alpha} }\left(p_{ij} - {1
\over 2}g_{ij}p\right) +\xi_{i;j} +\xi_{j;i},\label{4.25}
\end{equation}
so we define $N = \sqrt{T/4R^{\alpha}}$. The square-root identity
becomes
\begin{equation}
    gH = -p^{ij}p_{ij} + {1 \over 2}p^{2}  + gR^{\alpha}= 0,
\label{4.26}
\end{equation}
and we get the standard momentum constraint
\begin{equation}
 \sqrt{g}H^j = p^{ij}_{~~;i} = 0. \label{4.27}
\end{equation}
The dynamical equations are
\begin{eqnarray}
{\partial   p^{ij} \over \partial \lambda} = {\delta {\cal L}
\over \delta g_{ij}} =& -\sqrt{g}N R^{\alpha - 1}\left(\alpha
R^{ij} - g^{ij}R \right) - 2N\left(p^{im}p_m^{~j} - {1 \over
2}pp^{ij}\right) \nonumber \\
&+\alpha\sqrt{g}\left(\left[NR^{\alpha - 1} \right]^{;ij} -
g^{ij}\nabla^2NR^{\alpha - 1}\right) +{\cal L}_{\xi^i}p^{ij}.
\label{4.28}
\end{eqnarray}
We use (\ref{4.25}) and (\ref{4.28}) to evolve the constraints.
Once again, the Hamiltonian constraint does not propagate:
\begin{eqnarray}
{\partial \over \partial \lambda}\left[\sqrt{g}R ^{\alpha}- {1
\over \sqrt{g}}\left( p^{ab}p_{ab} - {1 \over
2}p^{2}\right)\right] &= 4\alpha R^{\alpha - 1}N^{;a}p^b_{~a;b} +
2\alpha R^{\alpha
- 1}Np^{ab}_{~~;ab}\nonumber \\
& + {1 \over 2} Np \left[\sqrt{g}R^{\alpha} - {1 \over
\sqrt{g}}\left( p^{ab}p_{ab} - {1 \over
2}p^{2}\right)\right]\nonumber \\& - 4 \alpha
N^{;a}\left(R^{\alpha - 1}\right)^{;b} p_{ab} - 2 \alpha N
\left(R^{\alpha - 1}\right)^{;ab}p_{ab} \nonumber \\&+ {\cal
L}_{N^i}\left[\sqrt{g}R^{\alpha} - {1 \over \sqrt{g}}\left(
p^{ab}p_{ab} - {1 \over 2}p^{2}\right)\right].
\label{4.29}\end{eqnarray} This does not vanish weakly for any
$\alpha$ except $\alpha = 1$. We get the extra constraint $R = $
constant. Conserving this gives yet another, unpleasant, equation.
It is difficult to conceive any solution of this system except
static flat space.

Another choice we tested was
\begin{equation}
P = C_1R^2 + C_2 R^{ab}R_{ab} + C_3 \nabla^2R, \label{4.30}
\end{equation}
where $C_1, C_2, C_3$ are arbitrary constants. If $g_{ij}$ is
taken to have dimensions (length)$^2$, then $R$ will be
(length)$^{-2}$. No scalar with dimensions (length)$^{-3}$ can be
constructed from $g_{ij}$ . The only geometric scalars that have
dimension (length)$^{-4}$ are the three in expression
(\ref{4.30}). The other two obvious candidates, the square of the
Riemann tensor and $R^{ij}_{~~;ij}$, need not appear. The
three-dimensional Riemann tensor can be written as a sum of the
Ricci tensor and $R$, and the divergence of the Ricci tensor can
be eliminated using the Bianchi identity.

We repeat the calculation, evolving the square-root constraint.
This leads to an explosion of unpleasant non-cancelling terms that
arise from the extra terms in $P$.

One soon sees that the same problems will arise for all possible
extra terms.  We conclude that BSW is the unique consistent
matter-free theory on superspace based on a `distance' of the form
(1.5).  We believe that this is a new result.  In many respects,
our calculations repeat those of HKT \cite{hkt1}.  The novelty is
our weaker assumption. The HKT assumptions are: 1) there is a
local Hamiltonian constraint, quadratic in the momenta; 2) there is a
local momentum constraint; 3) the Poisson bracket of these constraints
reflects embeddability. Our local square root is equivalent to 1); best
matching is equivalent to 2); constraint propagation on its own
replaces 3).  There is no need to presuppose spacetime. Already
latent in (1.5), it is laid bare by consistency.

\section{Scalar Field Interacting with Gravity}

There exist matter-free solutions of Einstein's equations on
$S^3$.  Thus, there is an emergent light-cone structure in the
3-space approach. Besides the pure-gravity light cone and
4-covariance (which we have recovered), the most basic
relativistic facts are the universality of free fall and
the universal light cone (all matter fields respecting
the gravity cone). If (\ref{1.6a}) is the basis of relativity,
both of these further features should be implied by it.

Let us see how a real scalar field $\phi$ can be introduced.
First, best matching essentially fixes the form in which $\phi$
enters the kinetic term $T$. For $\phi$ is `painted' onto the
3-geometries described by the 3-metrics $g_{ij}$, so that the
correction to its `naive' velocity $\partial\phi/\partial\lambda$,
like the correction $K\xi$ to the metric velocity $\partial
g_{ij}/\partial \lambda$ induced by (\ref{kf}), is predetermined.
It is the scalar product of $\xi$ with the spatial gradient of
$\phi$: the matter is `dragged along' with the geometry by the
diffeomorphisms. Technically, the correction term
$\phi_{;i}\xi^{i}$ is just the Lie derivative of $\phi$ along
$\xi$ just as $K\xi$ is the Lie derivative of $g_{ij}$. The
modified $T$ is
\begin{equation}
T = (g^{ac}g^{bd} - g^{ab}g^{cd}) \left[{\partial g_{ab} \over
\partial \lambda} - (K\xi)_{ab} \right]\left[{\partial g_{cd}
\over
\partial\lambda} - (K\xi)_{cd}\right] +  \left[{\partial \phi
\over
\partial \lambda} - \phi_{;i}\xi^i\right]^2. \label{5.1}
\end{equation}
As here, the coefficient of the scalar kinetic term can always be
set to 1 by absorbing a constant into $\phi$. The obvious
modifications to the potential $P$ are
\begin{equation}
 R \rightarrow  R   - {C\over
4}g^{ab}\phi_{,a}\phi_{,b} +\sum_n A_n \phi^n. \label{5.2}
\end{equation}

The first addition is the standard scalar-field term that gives
rise to wave propagation. It has the same dimensions,
(length)$^{-2}$, as $R$. If the constant $C \ne +1$, then $\phi$
will not have the same light cone as gravity and local Lorentz
invariance will be violated. The second addition is a general
polynomial non-derivative self-interaction term for $\phi$. For $n
= 2$ and $A_2 = m^2/4$, we get the standard mass term for $\phi$.
We need not demand that $n$ be an integer. We have dropped
$\Lambda$, but it can easily be restored. We include neither the
higher-order metric terms excluded in Sec. VI nor higher-order
metric--scalar interactions. We expect that these too can be
eliminated.

The metric momenta are
\begin{equation}
    p^{ij} = {\delta {\cal L} \over \delta\left({\partial g_{ij} \over
\partial \lambda}\right)} = \sqrt{g(R + U_{\phi}) \over T_g +
T_{\phi}}(g^{ic}g^{jd} - g^{ij}g^{cd}) \left[{\partial g_{cd}
\over \partial\lambda} - (K\xi)_{cd}\right], \label{5.3}
\end{equation}
where $U_{\phi}$ is the $\phi$ potential term. Inversion of
(\ref{5.3}) gives
\begin{equation}
{\partial g_{ij} \over \partial \lambda} = {2N \over\sqrt{g}}
\left(p_{ij}  - {1 \over 2}g_{ij}p\right) +\xi_{i;j}
+\xi_{j;i}.\label{5.4}
\end{equation}
where we define $2N = \sqrt{T_g + T_{\phi}/R + U_{\phi}}$. The
momentum conjugate to $\phi$ is
\begin{equation}
    \pi = {\delta {\cal L} \over \delta\left({\partial \phi \over
\partial \lambda}\right)} = \sqrt{g(R + U_{\phi}) \over T_g +
T_{\phi}} \left[{\partial \phi \over \partial\lambda} -
\phi_{,i}\xi^i\right]. \label{5.5}
\end{equation}
Its inversion gives
\begin{equation}
{\partial \phi \over \partial \lambda} = {2N\pi \over\sqrt{g}}
 +\phi_{,i}\xi^i.\label{5.6}
\end{equation}
The square-root identity becomes
\begin{equation}
    gH = -p^{ij}p_{ij} + {1 \over 2}p^{2} - \pi^2 + g(R +
U_{\phi})= 0, \label{5.7}
\end{equation}
and variation w.r.t. $\xi$ gives the momentum constraint
\begin{equation}
 \sqrt{g}H^j = p^{ij}_{~~;i} - {1 \over 2}\pi\phi^{;j}= 0. \label{5.8}
\end{equation}
The Euler--Lagrange equations for $g_{ij}$ and $\phi$ are
\begin{eqnarray}
{\partial   p^{ij} \over \partial \lambda} = {\delta {\cal L}
\over \delta g_{ij}} =& -\sqrt{g}N\left(R^{ij} - g^{ij}R \right) -
{2N \over \sqrt{g}}\left(p^{im}p_m^{~j} - {1 \over
2}pp^{ij}\right)  +\sqrt{g}\left(N^{;ij}  - g^{ij}\nabla^2N\right)
\nonumber \\ & + {\sqrt{g}CN \over 4}\nabla^i\phi\nabla^j\phi -
{\sqrt{g}C N \over 4}g^{ab}\phi_{,a}\phi_{,b}g^{ij} +
\sqrt{g}N\sum_n A_n \phi^n g^{ij}. +{\cal L}_{\xi^i}p^{ij}
\label{5.9}\\
{\partial   \pi \over \partial \lambda} = {\delta {\cal L} \over
\delta \phi} =& {\sqrt{g} C \over 2} (N\nabla^i\phi)_{;i}
+\sqrt{g}N \sum n A_n\phi^{n-1} + {\cal L}_{\xi^i}\pi.
\label{5.10}
\end{eqnarray}
We can now combine (\ref{5.6}) and (\ref{5.10}) to give
\begin{equation}
{\partial \over \partial \lambda}\left({\sqrt{g} \over N}
\left[{\partial \phi \over \partial\lambda} -
\phi_{,i}\xi^i\right]\right) - \sqrt{g} C (N\nabla^i\phi)_{;i} -
 2\sqrt{g}N\sum n A_n \phi^{n-1}
 - 2{\cal L}_{\xi^i}\pi= 0.
\label{5.11}
\end{equation}
If we set $\sqrt{g} = N = 1, n = 2, A_2 = m^2/4, \xi = 0$, this
reduces to $\partial^2\phi/\partial^2\lambda - C\nabla^2\phi - m^2
\phi = 0$, i.e., the wave equation for a scalar field with mass
$m$ and canonical speed $\sqrt{C}$.

The new Hamiltonian and momentum constraints (\ref{5.7}) and
(\ref{5.8}) are the 00 and 0$i$ Einstein field equations. Note
that $\pi\phi^{;j}$ from (\ref{5.8}) completes the square of
$\pi^2 + 1/4 (\nabla\phi)^2$ from (\ref{5.7}). The 1/2 in
(\ref{5.8}) arises because the Hamiltonian constraint has
$16\pi\rho$, while the momentum constraint has $8 \pi J^i$. These
are obtained because of our best-matching `dragging' of $\phi$,
which leads to minimal coupling and the equivalence principle.
Note that the divergence $ p^{ij}_{~~;i}$ of the gravitational
momenta no longer vanishes but equals the $\phi$ `current'
${1\over 2}\phi\pi^{;j}$ (the $\phi$ momentum density), while the
Hamiltonian constraint picks up (twice) the energy density.

We now come to the next constraint-propagation result. The
momentum constraint propagates, as always, and the polynomial
self-interaction terms (including the mass term) in the quadratic
constraint cause no difficulty; the extra terms in its $\lambda$
derivative cancel. But the coefficient $C$ produces non-vanishing
terms. We have
\begin{eqnarray}
&{\partial \over \partial \lambda}\left[\sqrt{g}\left(R  - {C\over
4}g^{ab}\phi_{,a}\phi_{,b} +\sum_n A_n \phi^n\right) - {1 \over
\sqrt{g}}\left( p^{ab}p_{ab} - {1 \over 2}p^{2} + \pi^2\right)
\right] \nonumber\\ & = 2N^{;a}\left(2p^b_{~a;b} - \pi
\phi_{;a}\right) + N\left(2p^{ab}_{~~;a} - \pi
\phi^{;b}\right)_{;b} \nonumber \\ &+{1 \over 2}Np
\left[\sqrt{g}\left(R - {C\over 4}g^{ab}\phi_{,a}\phi_{,b} +\sum_n
A_n \phi^n\right) - {1 \over \sqrt{g}}\left( p^{ab}p_{ab} - {1
\over 2}p^{2} + \pi^2\right)\right]\nonumber \\ &+ {\cal
L}_{N^i}\left[\sqrt{g}\left(R - {C\over 4}g^{ab}\phi_{,a}\phi_{,b}
+\sum_n A_n \phi^n\right)- {1 \over \sqrt{g}}\left( p^{ab}p_{ab} -
{1 \over 2}p^{2} + \pi^2\right)\right]\nonumber \\
& +(1 - C)N\pi\nabla^2\phi +(2 - 2C)N^{;i}\pi\phi_{;i} +(1 -
C)N\pi^{;i}\phi_{;i}. \label{5.12}\end{eqnarray}

Although most terms in (\ref{5.12}) vanish weakly, the last three,
all proportional to $1 - C$, do not. They generate a secondary
constraint. This forms a second-class constraint set with the
Hamiltonian and so generates yet another constraint. In fact, the
Hamiltonian constraint will not propagate unless $C = 1$. But this
means that $\phi$ is forced to respect the metric light cone.

Moreover, the mechanism that enforces this will generate the
universal metric--matter light cone -- for bosons at least, we
have not yet considered fermions. The mechanism has
several hinges, but the linkage is unbreakable. The key is the
presence of the scalar curvature $R$ in the square-root constraint.
The
$\lambda$ derivative of this constraint therefore contains
$\partial R/\partial\lambda$. Now purely by kinematics
\begin{equation}
{\partial R\over \partial \lambda} = \left[{\partial g_{ij}\over
\partial
\lambda}\right]^{;ij} - \nabla^2\left[g^{ij}{\partial g_{ij}\over
\partial \lambda}\right] - R^{ij}{\partial g_{ij}\over \partial
\lambda} = \left[{2N \over\sqrt{g}}p^{ij}\right]_{;ij} +
\dots,\label{7.13a}
\end{equation}
where we use Eq.(\ref{5.4}) to replace the time derivative of $g$
with the momentum. The final expression in
Eq.(\ref{7.13a}) will contain terms with $p^{ij}_{;i}$.  We now
subtract
$\pi\phi^{;i}$ to generate the momentum constraint and obtain the
first two terms on the right hand side of Eq.(\ref{5.12}). This is
the first hinge. The extra
$\pi\phi^{;i}$ terms now appear as the
$C$-independent parts of the last three terms in Eq.(\ref{5.12}).
The $\pi\phi^{;i}$ terms with the factor $C$ arise from two
different sources. The $(\nabla\phi)^2$ term in the
Hamiltonian constraint has an explicit $C$ in front of it.
The time derivative of this term, combined with
Eq.(\ref{5.6}) gives the term in the middle of the three.
The other two terms arise from the time derivative of the
$\pi^2$ in the square root constraint. To evaluate this we
use the Euler-Lagrange equation for $(\phi, \pi)$,
Eq.(\ref{5.10}); the first term in this has a $C$ in it.
This gives the first and third terms. This is the second hinge.

Therefore, the
constraint-propagation condition contains identical terms with and
without the coefficient. This is the final hinge in the light-cone
generating mechanism, for it makes it inevitable that constraint
propagation will enforce $C = 1$. The universality of the mechanism
is ensured by the universal nature of the momentum constraint: any
field of whatever tensor rank whose velocity appears in the
Lagrangian with best-matched correction will be represented as a
source term in the momentum constraint in a form solely determined
by its Lie derivative. We shall see in the next section that this
mechanism has even more implications.

For the scalar field, we have thus derived the correct light-cone
behaviour of Lorentz-invariant field theory and can see that this
will hold uniformly. We believe that this is a new result. We can
also show that a derivative coupling term in (\ref{5.2}) cannot be
included consistently in the potential of our Lagrangian.

It turns out that the fact that the `mass' term in the scalar field
can be chosen quite arbitrarily, as we have shown above, is part of a
wider freedom. It turns out that a large class of what might be called
`dilatonic' theories can be written selfconsistently in BSW form. Let
us consider a theory which has three constituents. One is the
gravitational field, represented by a three-metric, $g_{ij}$; one
is a massless scalar field, $\phi$ (the dilaton); and the third
is a `massive' scalar field, $\chi$. The key point is that we
assume that the `mass' of $\chi$ is some arbitrary function of
$\phi$, $m^2 = f(\phi)$. We further assume that both scalar fields
are minimally coupled to the metric and that the kinetic energy takes
the simplest possible form.

This means that we assume the potential term is
\begin{equation}
 P =  R   - {1\over
4}g^{ab}\phi_{,a}\phi_{,b} - {1\over
4}\left(g^{ab}\chi_{,a}\chi_{,b} - f(\phi)\chi^2\right), \label{D1}
\end{equation}
and the kinetic term is
\begin{equation}
T = (g^{ac}g^{bd} - g^{ab}g^{cd}) \left[{\partial g_{ab} \over
\partial \lambda} - (K\xi)_{ab} \right]\left[{\partial g_{cd}
\over
\partial\lambda} - (K\xi)_{cd}\right] +  \left[{\partial \phi
\over
\partial \lambda} - \phi_{;i}\xi^i\right]^2 +  \left[{\partial \chi
\over
\partial \lambda} - \chi_{;i}\xi^i\right]^2. \label{D2}
\end{equation}
The square-root identity becomes
\begin{equation}
    gH = -p^{ij}p_{ij} + {1 \over 2}p^{2} - \pi_{\phi}^2 -
\pi_{\chi}^2 + g(R + U_{\phi} +U_{\chi})= 0, \label{D3}
\end{equation}
and variation w.r.t. $\xi$ gives the momentum constraint
\begin{equation}
 \sqrt{g}H^j = p^{ij}_{~~;i} - {1 \over 2}\pi_{\phi}\phi^{;j} - {1
\over 2}\pi_{\chi}\chi^{;j}= 0.
\label{D4}
\end{equation}
We need to consider the propogation of the square-root constraint. We
know that most of the terms will take care of themselves and we really
need only focus on the derivative of $f(\phi)$. One place is in
the Euler--Lagrange equation for $\phi$
\begin{equation}
{\partial   \pi_{\phi} \over \partial \lambda} = {\delta {\cal L} \over
\delta \phi} = {\sqrt{g}  \over 2} (N\nabla^i\phi)_{;i}
+{\sqrt{g}N \over 4}f'(\phi)\chi^2 + {\cal L}_{\xi^i}\pi_{\phi}.
\label{D5}
\end{equation}
This will feed into
\begin{equation}
-{\partial   \pi_{\phi}^2 \over \partial \lambda} = -{\sqrt{g}N
\pi_{\phi}\over 2}f'(\phi)\chi^2 + \dots.
\label{D6}
\end{equation}
The other place is in the time derivative of $U_{\chi}$ where we get
\begin{equation}
{\partial   f(\phi) \over \partial \lambda}{g\chi^2 \over 4} =
{gf'(\phi)\chi^2 \over 4}{\partial   \phi \over \partial \lambda} =
{gf'(\phi)\chi^2 \over 4}{2N\pi_{\phi} \over \sqrt{g}} = {\sqrt{g}N
\pi_{\phi}\over 2}f'(\phi)\chi^2, \label{D7}
\end{equation}
where we used Eq.(\ref{5.6}) to replace the time derivative of $\phi$
with its momentum. The two terms obviously cancel. It is clear that we
can replace the mass term in $\chi$ with a polynomial and have each
coefficient be a different function of $\phi$.

Possibly the best known dilaton theory is Brans--Dicke theory
\cite{B-D}. It is interesting to see how this can be written in BSW
form. Let us start with the simplest possible local square root form
with just gravity and a minimally coupled massless scalar field. This
means that we choose the potential and kinetic terms as in Eqns.
(\ref{D1}) and (\ref{D2}) with $\chi \equiv 0$. We now perform a
`point' transformation and change variables to $(\gamma, \Phi)$ via:
\begin{equation}
g_{ab} = \Phi \gamma_{ab}, \hskip 1cm \phi = -\sqrt{4\omega +
6}\log\Phi. \label{B-D1}
\end{equation}
The action changes to
\begin{eqnarray}
&\sqrt{g}\sqrt{P}\sqrt{T} = \sqrt{\gamma}\sqrt{\Phi^2 R -
\omega \gamma^{ij}\partial_i\Phi\partial_j\Phi  -
2\Phi\nabla^2\Phi}\times\nonumber\\
&\sqrt{(\gamma^{ac}\gamma^{bd} - \gamma^{ab}\gamma^{cd})
\left[{\partial \gamma_{ab} \over \partial \lambda} -
(K{\bar\xi})_{ab} \right]\left[{\partial \gamma_{cd} \over
\partial\lambda} - (K{\bar\xi})_{cd}\right] - {4 \over \Phi}\gamma^{cd}
\left[{\partial \gamma_{cd} \over
\partial\lambda} - (K{\bar\xi})_{cd}\right]\left[{\partial
\Phi \over
\partial \lambda} - \Phi_{;i}\bar\xi^i\right] +  {4\omega \over
\Phi^2}\left[{\partial \Phi\over
\partial \lambda} - \Phi_{;i}\bar\xi^i\right]^2},\label{B-D}
\end{eqnarray}
where $\bar\xi^i = \xi^i$ and $\bar\xi_i = \Phi^{-1}\xi_i$.
We vary this w.r.t. $\partial \gamma_{cd}/
\partial\lambda$ and $\partial
\Phi/\partial \lambda$ to get the momenta conjugate to
$\gamma_{cd}$ and $\Phi$. These are
\begin{equation}
p^{cd}_\gamma = \sqrt{\gamma P\over
T}\left\{(\gamma^{ac}\gamma^{bd} -
\gamma^{ab}\gamma^{cd})\left[{\partial \gamma_{ab} \over \partial
\lambda} - (K{\bar\xi})_{ab} \right] - {2 \over
\Phi}\gamma^{cd}\left[{\partial \Phi \over \partial \lambda} -
\Phi_{;i}\bar\xi^i\right]\right\}
\end{equation}
and
\begin{equation}
\pi_{\Phi} = \sqrt{\gamma P\over T}\left\{{4\omega \over
\Phi^2}\left[{\partial \Phi\over
\partial \lambda} - \Phi_{;i}\bar\xi^i\right] - {2 \over \Phi}\gamma^{cd}
\left[{\partial \gamma_{cd} \over
\partial\lambda} - (K{\bar\xi})_{cd}\right]\right\}.
\end{equation}

Then the square-root constraint simply becomes the standard 3+1
Brans--Dicke Hamiltonian constraint and similarly for the momentum
constraint.
The square-root identity becomes
\begin{equation}
    \gamma H = -p^{ij}p_{ij} + {1 \over 2}p^{2} - {1 \over
(4\omega + 6)}\left(p - \Phi\pi_{\Phi}\right)^2
 + \Phi^2 R -
\omega \gamma^{ij}\partial_i\Phi\partial_j\Phi  -
2\Phi\nabla^2\Phi= 0, \label{B-D3}
\end{equation}
and variation w.r.t. $\xi$ gives the momentum constraint
\begin{equation}
 \sqrt{\gamma}H^j = p^{ij}_{\gamma;i} - {1 \over 2}\pi_{\Phi}\Phi^{;j}
= 0.\label{B-D4}
\end{equation}
This constraint algebra will close because it is just a transformed
version of a minimally coupled scalar field.

 There is thus a one-to-one correspondence between
solutions of the Einstein equations with a minimally coupled
massless scalar field and the solutions of the `vacuum'
Brans--Dicke equations. The true difference between the two
theories is whether the Einstein metric $g$ or the Brans--Dicke
metric $\gamma$ determines the geodesics. One way to test this is
to add a second minimally-coupled scalar field.

Therefore, consider the Brans--Dicke action (\ref{B-D}) and
minimally couple a second, constant mass, scalar field $\chi$ to it. To
do this, we add  $\Delta P = -\Phi^2{1 \over
4}(\gamma^{ij}\chi_{,i}\chi_{,j} - m^2\chi^2)$ to the potential term
and add
$\Delta T = [\partial\chi/\partial\lambda -
\chi_{;i}\bar\xi^i]^2$ to the kinetic term. Rather than computing in
the Brans--Dicke frame it is easier to see what is happening in the
Einstein frame.  Let us now  undo
the transformation (\ref{B-D1}). The Brans--Dicke field translates into
Einstein gravity with a minimally-coupled scalar field. The kinetic
 term is of the standard form,
\begin{equation}
T = (g^{ac}g^{bd} - g^{ab}g^{cd}) \left[{\partial g_{ab} \over
\partial \lambda} - (K\xi)_{ab} \right]\left[{\partial g_{cd}
\over
\partial\lambda} - (K\xi)_{cd}\right] +  \left[{\partial \phi
\over
\partial \lambda} - \phi_{;i}\xi^i\right]^2 +  \left[{\partial \chi
\over
\partial \lambda} - \chi_{;i}\xi^i\right]^2, \label{B-D5}
\end{equation}
 the additional potential term becomes
\begin{equation}
 P =  R   - {1\over
4}g^{ab}\phi_{,a}\phi_{,b} - {1\over
4}\left(g^{ab}\chi_{,a}\chi_{,b} -
e^{\phi\over\sqrt{4\omega + 6}}\chi^2\right).
\label{B-D6}
\end{equation}
This is a special case of the dilaton theories we discussed
earlier. Therefore the constraint algebra will close both in the
Einstein frame and in the Brans--Dicke frame. Therefore Brans--Dicke
fits very naturally into the BSW framework.

The conclusion of this section must be that, when we consider scalar
fields coupled to gravity and insist on a local square root action, we
recover causality. The characteristic speeds of the scalar fields must
coincide with that of gravity. Other than that, there is a large
freedom in the detailed form of the scalar fields and in the way they
interact. In the next section we deal with vector fields. There we
discover that the detailed form of the vector field is rigidly
prescribed. The only theory that fits our framework is massless
electrodynamics, we recover the Maxwell equations unambiguously.

\section{Three-Vector Field Interacting with Gravity}

Since the kinetic and potential terms of different fields are
simply added separately to the potential and $T$ and do not mix
unless an interaction between them is introduced explicitly, we
can treat different fields (scalar, 3-vector, 3-spinor)
separately. We now consider a 3-vector field $A_a$. We use the
covariant $A_a$ as the independent object, matching our use of the
covariant metric $g_{ab}$ and covariant shift $\xi_a$.

The correction to the 3-vector velocity induced by the
diffeomorphism-generating field $\xi$ to implement best matching
is as unambiguous as it is for the scalar field.  It is the Lie
derivative of $A_a$ with respect to $\xi$:
\begin{equation}
{\partial A_a \over \partial \lambda} \rightarrow {\partial A_a
\over
\partial \lambda} - {\cal L}_{\xi}A_a = {\partial A_a \over \partial
\lambda} - g^{bc}A_{a;b}\xi_c - g^{bc}\xi_{b;a}A_c. \label{6.1}
\end{equation}
Therefore we add a term
\begin{equation}
T_A = g^{ad}\left({\partial A_a \over \partial \lambda} -
g^{bc}A_{a;b}\xi_c - g^{bc}\xi_{b;a}A_c\right) \left({\partial A_d
\over \partial \lambda} - g^{bc}A_{d;b}\xi_c -
g^{bc}\xi_{b;d}A_c\right)
\end{equation}
to the metric kinetic energy.

The additions $U_A$ to the potential are equally obvious:
\begin{eqnarray}
    U_A =&  C_1A_{a;b}A^{a;b} + C_2A_{a;b}A^{b;a} + C_3A^a_{;a}A^b_{;b} +
\sum_kB_k(A^aA_a)^k \nonumber \\=& \left(C_1g^{ab}g^{cd} +
C_2g^{ad}g^{bc} + C_3g^{ac}g^{bd}\right)A_{a;c}A_{b;d} +
\sum_kB_k(g^{ab}A_aA_b)^k.
\end{eqnarray}
Hence we begin with the modified BSW action (which we call $A_A$)
\begin{equation}
A_A = \int d\lambda \int \sqrt{g} \sqrt{R + U_A} \sqrt{T_g + T_A}
d^3x,
\end{equation}
and vary it to get
\begin{equation}
    p^{ij} = {\delta {\cal L} \over \delta\left({\partial g_{ij} \over
\partial \lambda}\right)} = \sqrt{g(R + U_A) \over T_g +
T_A}(g^{ic}g^{jd} - g^{ij}g^{cd}) \left[{\partial g_{cd} \over
\partial\lambda} - (K\xi)_{cd}\right]. \label{6.5}
\end{equation}
This can be inverted to give
\begin{equation}
{\partial g_{ij} \over \partial \lambda} = {2N \over\sqrt{g}}
\left(p_{ij}  - {1 \over 2}g_{ij}p\right) +\xi_{i;j}
+\xi_{j;i}.\label{6.6}
\end{equation}
where we define $2N = \sqrt{T_g + T_A/R + U_A}$. The momentum
conjugate to $A_a$ is
\begin{equation}
    \pi^a = {\delta {\cal L} \over \delta\left({\partial A_a \over
\partial \lambda}\right)} = \sqrt{g(R + U_A) \over T_g +
T_A}g^{ab} \left[{\partial A_b \over \partial\lambda} - {\cal
L}_{\xi}A_b\right]. \label{6.7}
\end{equation}
This also can be inverted to give
\begin{equation}
{\partial A_b \over \partial \lambda} = {2N\pi_b \over\sqrt{g}}
 + {\cal L}_{\xi}A_b.\label{6.8}
\end{equation}

The square-root identity becomes
\begin{equation}
    gH = -p^{ij}p_{ij} + {1 \over 2}p^{2} - \pi^a\pi_a + g(R +
U_A)= 0, \label{6.9}
\end{equation}
and variation w.r.t. $\xi$ gives the momentum constraint
\begin{equation}
 \sqrt{g}H^j = p^{ij}_{~~;i} - {1 \over 2}\left(\pi^cA_c^{~;j} -
\left[\pi^bA^j\right]_{;b}\right) = p^{ij}_{~~;i} -{1 \over
2}\left(\pi^c\left[A_c^{~;j} - A^j_{~;c}\right] -
\pi^b_{;b}A^j\right) = 0. \label{6.10}
\end{equation}
The Euler--Lagrange evolution equations for $g_{ij}$ and $A_a$ are
\begin{eqnarray}
{\partial   p^{ij} \over \partial \lambda} = {\delta {\cal L}
\over \delta g_{ij}} =& -\sqrt{g}N\left(R^{ij} - g^{ij}R \right) -
{2N \over \sqrt{g}}\left(p^{im}p_m^{~j} - {1 \over
2}pp^{ij}\right)  +\sqrt{g}\left(N^{;ij}  -
g^{ij}\nabla^2N\right) \nonumber \\
& - \sqrt{g} N \left[C_1\left(A^{i;a}A^j_{~;a} +
A^{a;i}A_a^{~;j}\right) +C_2\left(A^{i;a}A_a^{~;j} +
A^{a;i}A^j_{~;a}\right) +C_3A^a_{~;a}\left(A^{i;j} +
A^{j;i}\right)\right] \nonumber \\&
 + \sqrt{g} N \left[C_1A_{a;b}A^{a;b} +
 C_2A_{a;b}A^{b;a} + C_3A^a_{;a}A^b_{;b}\right]g^{ij} -
N{\pi^i\pi^j \over \sqrt{g}} \nonumber\\&
+\sqrt{g}  \left[C_1 + C_2\right]\left[NA^{(i}A^{j);c} + NA^{(i}A^{|c|;j)} -
NA^cA^{(i;j)}\right]_{;c}  + \sqrt{g} C_3\left[2NA^d_{;d}g^{(i|c|}A^{j)} -
NA^d_{;d}g^{ij}A^c\right]_{;c} \nonumber\\&-
\sqrt{g}N\sum_kkB_k (A^aA_a)^{k - 1}A^iA^j +
\sqrt{g}N\sum_kB_k(A^aA_a)^kg^{ij} +{\cal L}_{\xi^i}p^{ij}
\label{6.11}\\
{\partial   \pi^i \over \partial \lambda} = {\delta {\cal L} \over
\delta A_i} =& -2\sqrt{g} \left[C_1\left(NA^i_{~;b}\right)^{;b}
+C_2\left(NA_b^{~;i}\right)^{;b} +
C_3\left(NA_b^{~;b}\right)^{;a}\right]
 +2\sqrt{g}N \sum k B_k(A^aA_a)^{k - 1}A^i + {\cal
L}_{\xi}\pi^i. \label{6.12}
\end{eqnarray}

We now check for propagation of the constraints (\ref{6.9}) and
(\ref{6.10}) under the evolution by (\ref{6.6}), (\ref{6.8}),
(\ref{6.11}) and (\ref{6.12}).
    As expected, the momentum constraint propagates.  In the Hamiltonian
constraint, the simple self-interaction terms (with coefficients
$B_k$) with no derivatives of either $A_a$ or $g_{ij}$ give no
problem at this stage. However, $\partial H/
\partial \lambda$ contains terms that are not proportional to
the constraints. The full expression is
\begin{eqnarray}
&{\partial \over \partial \lambda}\left[\sqrt{g}\left(R   +
U_A\right) - {1 \over \sqrt{g}}\left( p^{ab}p_{ab} - {1 \over
2}p^{2} + \pi^a\pi_a\right) \right] = \nonumber\\&
2N^{;a}\left(2p^b_{~a;b} - \pi^c\left[A_c^{~;a} - A^a_{~;c}\right]
+ \pi^b_{;b}A^a\right) + N\left(2p^{ab}_{~~;a}
-\pi^c\left[A_c^{~;b} - A^b_{~;c}\right] +
\pi^a_{;a}A^b\right)_{;b} \nonumber \\ &+{1 \over 2}Np
\left[\sqrt{g}\left(R + U_A\right) - {1 \over \sqrt{g}}\left(
p^{ab}p_{ab} - {1 \over 2}p^{2} + \pi^a\pi_a\right)\right]
\nonumber
\\ &+ {\cal L}_{N^i}\left[\sqrt{g}\left(R + U_A\right)- {1 \over
\sqrt{g}}\left( p^{ab}p_{ab} - {1
\over 2}p^{2} + \pi^a\pi_a\right)\right]\nonumber \\
& +{4C_1 + 1 \over N}\left(N^2\pi^aA_a^{~;b}\right)_{;b} + {4C_2 -
1 \over N}\left(N^2\pi^aA^b_{~;a}\right)_{;b} - {1 \over
N}\left(N^2\pi^a_{~;a}A^b\right)_{;b} + {4C_3 \over N}\left(N^2
\pi^aA^b_{~;b}\right)_{;a}\nonumber \\
& -2{C_1 + C_2\over N}\{2[N^2A_{(b;d)}A^f(p^b_f - {1 \over
2}p\delta^b_f)]^{;d} - [N^2A_{(b;d)}A^f(p^{bd} - {1 \over 2}pg^{bd})]_{;f}\}
\nonumber \\
&- 2{C_3\over N}\{2[N^2A^a_{;a}g_{bd}A^f(p^b_f - {1 \over
2}p\delta^b_f)]^{;d} - [N^2A^a_{;a}g_{bd}A^f(p^{bd} - {1 \over
2}pg^{bd})]_{;f}\} .
\label{6.13}\end{eqnarray}
The last six terms do not vanish
weakly. They have to vanish, otherwise we get a family of
secondary constraints which eliminates all the degrees of freedom.
We can (and must!) eliminate most of them by choosing $C_1 = -C_2
= -1/4, C_3 = 0$ (which means that $U_A = -({\rm curl}A)^2/4$).
This still leaves us with $\pi^b_{~;b} = 0$ as a new, secondary, constraint
which we cannot eliminate. This is none other than the Gauss constraint.

Propagation of this new constraint requires
\begin{equation}
{\partial \over \partial \lambda}\pi^b_{~;b} = +2\sqrt{g}\left[N
\sum k B_k(A^aA_a)^{k - 1}A^i\right]_{;i} + {\cal
L}_{\xi}\pi^i_{~;i}. \label{6.14}
\end{equation}
The only way to ensure propagation is to set $B_k = 0~\forall~k$.
Therefore, the previously  unproblematic undifferentiated
potential terms are incompatible with the extra constraint. In our
scheme, a scalar field can have mass (or power-law nonlinear
coupling, say $\phi^4$) but a 3-vector field cannot. Only the
Maxwell 3-vector field minimally coupled to gravity and locked to
its light cone survives.

Since only ${\rm curl}$ $A$ appears in the potential, addition of
a time-independent scalar to $A_{a}$, $A_a \rightarrow A_a +
\partial_a\Lambda$, does not change the action. This is the
analogue of a time-independent 3-coordinate transformation. If,
more appropriately, we take a time-dependent $\Lambda$, we need to
add the extra term $-\partial_a\Phi$ to (\ref{6.1}). The role of
$\Phi$ is to absorb the $\partial \Lambda /\partial\lambda$ that
arises from the $\partial A/\partial \lambda$ term. The $\xi$
terms play just the same role in compensating time-dependent
coordinate transformations.

It is now is easy to build the divergence constraint $\pi^b_{~;b}
= 0$ into the action.  We simply regard $\Phi$ as an independent
variable and vary w.r.t. it, just as with $\xi$. Thus we regard
$A_a$ as a gauge field, so that $\pi^b_{~;b} = 0$ becomes the
Gauss constraint. The full correction to (\ref{6.1}) is now
\begin{equation}
{\partial A_a \over \partial \lambda} \rightarrow {\partial A_a
\over
\partial \lambda} - \Phi_{,a}- {\cal L}_{\xi}A_a =
{\partial A_a\over\partial \lambda} - g^{bc}A_{a;b}\xi_c -
g^{bc}\xi_{b;a}A_c - \Phi_{,a}, \label{6.15}
\end{equation}
and we minimize w.r.t. $\Phi$ in exactly the same way as w.r.t.
$\xi$. This `gauge best matching' yields the Gauss constraint.
Thus, the only way we have found to propagate the constraints of a
metric--vector BSW-type action is to take
\begin{eqnarray}
U_A &=&{1 \over 4}\left(A^{a;b}A_{b;a} - A^{a;b}A_{a;b}\right) =
-{1 \over
4}\left({\rm curl}A\right)^2 \nonumber\\
T = T_g + T_A &=& (g^{ac}g^{bd} - g^{ab}g^{cd}) \left[{\partial
g_{ab} \over \partial \lambda} - (K\xi)_{ab}
\right]\left[{\partial g_{cd} \over
\partial\lambda} - (K\xi)_{cd}\right]\nonumber\\ & &+g^{ad}\left[{\partial
A_a \over
\partial
\lambda} - g^{bc}A_{a;b}\xi_c - g^{bc}\xi_{b;a}A_c -
\Phi_{,a}\right] \left[{\partial A_d \over
\partial
\lambda} - g^{bc}A_{d;b}\xi_c - g^{bc}\xi_{b;d}A_c -
\Phi_{,d}\right], \label{6.16}\end{eqnarray} which is simply
Einstein--Maxwell. There is, however, a technical point to note
here. The Lie-corrected velocity that appears in the square
parentheses in the final line of (\ref{6.16}) does not appear in the
form generally encountered in the literature. However, we can rewrite
\begin{eqnarray}
{\partial A_a \over \partial \lambda} - g^{bc}A_{a;b}\xi_c -
g^{bc}\xi_{b;a}A_c - \Phi_{,a} &=&
{\partial A_a \over \partial \lambda} - g^{bc}A_{a;b}\xi_c
+ g^{bc}A_{b;a}\xi_c - g^{bc}A_{b;a}\xi_c -
g^{bc}\xi_{b;a}A_c - \Phi_{,a}\nonumber\\ &=& {\partial A_a \over
\partial
\lambda}
 - g^{bc}A_{[a;b]}\xi_c -
[\xi^cA_c]_{;a} - \Phi_{,a} = {\partial A_a \over \partial \lambda}
 - g^{bc}A_{[a;b]}\xi_c -
  \Psi_{,a}
. \label{ambiguity} \end{eqnarray}

Here, we have replaced the gauge variable $\Phi$ by a new scalar
variable $\Psi = \Phi + \xi^cA_c$.
This means that in (\ref{ambiguity}) we now have a term linear
in the undifferentiated shift which is just the
familiar minimally-coupled Maxwell field tensor. The residual
scalar part that results from this manipulation gets absorbed in
the original gauge variable $\Phi$, which is the reason why we
denote this new composite scalar by $\Psi$. Two points are worth
mentioning here: 1) The Lie derivative of a 3-vector field becomes
ambiguous if the field is gauged, as is reflected in the
heterogeneous gauge--diffeomorphism nature of $\Psi$; 2) we are
obtaining highly sophisticated theories -- Einsteinian gravity and
Maxwell minimally coupled to gravity -- essentially uniquely out
of the 3-space approach but always only `by the skin of our
teeth'.

It is illuminating to consider why a massive vector field, which
is a perfectly good generally covariant theory,
is not allowed in the 3-space approach. As Giulini pointed out to
us \cite{giuliniPC}, the ADM decomposition for the massive vector
field is quite different from the massless case.
While the momentum can be written as
\begin{equation}
\pi_a = {\sqrt{g} \over N}\left[{\partial A_a \over \partial t}
 - F_{ab}N^b -
  A_{0,a}\right]\label{mvf1}
\end{equation} which looks identical to the last term in
Eq.(\ref{ambiguity}), there is a major difference in that, while
$\Psi$ is a true gauge degree of freedom and can be chosen
arbitrarily, $A_0$ in the massive vector field case is fixed. There
exists a primary constraint that the momentum conjugate to $A_0$,
$\pi^0$, vanishes. The conservation of this constraint leads to
\begin{equation}
\pi^i_{;i} + {\sqrt{g} \over N}m^2\left[A_0 - N^aA_a\right] = 0.
\label{mvf2}
\end{equation}

This is quite different from the massless case, for which, of
course, $m=0$, so that the second term is zero and we get a constraint
that is homogeneous (and linear) in the canonical momenta and contains
neither $A_{0}$ nor the shift. It is this restriction to
the canonical momenta that limits the physical degrees of freedom
and indicates that the action is defined on curves in the space that is
 the product of superspace and the transverse degrees of freedom of the
vector field. This matches our ideal of a geodesic-type theory. In the
massive case, the restriction is lifted since the part homogeneous in
the canonical momenta is now equal to a term containing what were
previously purely gauge variables. In whatever way one attempts to
interpret this extra term, it is clear that the massive vector field
does not belong to the class of theories
in which best matching with a local square root is applied to
{\it{bona fide}} three-dimensional geometrical objects.

Thus, we find that a non-gauge vector field cannot be coupled in
any simple manner to the BSW action.  For the reasons to be
explained in the conclusions, we think it would be premature to
attempt a rigorous no-go theorem, but we feel the provisional
result is already remarkable and even hints at a partial
unification of gravity and electromagnetism.  The fact is that in
pure geometrodynamics the BSW action is essentially unique, and we
have found only one vector field that couples to it: Maxwell.
Since our approach exploits 3-geometry to the maximal extent
possible but nothing else, we can say that Maxwellian theory is
uniquely inherent in Riemannian 3-geometries. We have already noted that,
within our scheme, the attempt to couple a single 3-vector field to
 scalar fields leads to the standard U(1) gauge coupling \cite{BFOM},
and the attempt to let a collection of 3-vector fields interact among
themselves leads to Yang-Mills gauge theory \cite{AB}.

Our result also gives further insight into the origin of Lorentz
invariance.  We begin with the field $A_a$, which is as
unashamedly three dimensional as the 3-metric. How does the full
panoply of the 4-potential $A_{\nu}, \nu = 1,2,3,4,$ and the
electromagnetic field tensor $F_{\mu\nu}$ arise?  The answer is
that the extra (`time') elements arise from the combined effect of
the Lie-derivative best-matching correction to the `bare'
3-velocity $\partial A_a/\partial\lambda$ and from having to
propagate the new square-root constraint. Best matching and the
local square root do it all. It is particularly striking that the
universal light cone and gauge theory arise from one and the same
Lie-derivative correction mechanism. For the scalar field, the
terms generated by the Lie correction in the momentum constraint
fix one constant in the constraint-propagation condition and lock
it to the gravity light cone. For the 3-vector field, they fix
three coefficients, with the same effect, and impose the Gauss
constraint. So, in a way, light cone and gauge are the same thing.
Both derive from best-matched gravity.

Some results of this section are only partly new, since Teitelboim
\cite{t} showed that his postulates (see the end of Sec. VI)
enforce gauge coupling. As in the case of the HKT result
\cite{hkt1}, we obtain his result with a significantly weaker
assumption, and we also obtain the light cone and reveal
its intimate connection to gauge theory.

\section{Concluding Remarks}

Hitherto it has always seemed that four-dimensional general
spacetime covariance is the very essence of GR. Many physicists
have expressed strong reservations about the 3+1 Hamiltonian
formalism.  It is held to be against the spirit of general
covariance and incapable of encompassing the wide range of
topologies allowed by GR.  The restriction to globally hyperbolic
spacetimes -- a necessary condition for the Hamiltonian treatment
-- is often severely criticized.

We believe that the present work, if it can be successfully
extended to the fermionic sector and withstands critical peer
review, puts these issues in a different light. The fact is that,
in a choice between two different theoretical schemes, there must
always be a preference (in the absence of compelling experimental
evidence) for the one that is more restrictive and, hence, makes
stronger predictions. There are two respects in which the 3-space
approach gives tighter predictions than the Einstein--Minkowski
approach: it rules out many fields (the massive vector field
 for example) and, within the possibilities that
remain, rules out many solutions, e.g., solutions with closed
timelike curves, which are not globally hyperbolic. These are
potential benefits.

Here we should also mention the possibility of advancing to
genuinely new physics. Our present paper does not actually predict
any new theory. It merely slims down a class of theories long
known to exist (all generally covariant theories) and, within the
restricted class, rules out exotic scenarios (time travel for
example). However, given that the treatment of fermions in curved
space is so delicate, we do not rule out the possibility that the
extension to the fermionic sector (if it succeeds) will bring
further insights. There is also the possibility \cite{bom},
already mentioned in the paper, of extension of the idea of best
matching to conformal superspace (CS).  As yet, we have performed
this extension only for the matter-free case, but the results so
far obtained are promising. They suggest the existence of a theory
virtually identical to GR except for the elimination of scale as a
dynamical variable.

If a full theory can be developed on CS, the implications are far
reaching. In the truly scale-invariant theory that will result,
the Hubble red shift cannot be explained by the `stretching of
space', as it is at present. For the theory on CS is designed
precisely to eliminate such `stretching' as a degree of freedom.
According to the present standard model, the universe is both
expanding and simultaneously changing its `shape', i.e., becoming
more `clumpy'. The expansion is used to explain the Hubble red
shift. In a scale-invariant theory, only change of shape is
physically meaningful, and it must explain the Hubble law though a
gravitational red shift induced by `clumping' {\cite{barb},
\cite{edge}). Thus, the 3-space approach does have the potential
to lead to very different cosmological predictions.

Finally, it is worth mentioning that the local square root creates
a theory that seems to be maximally sensitive to all properties of
Riemannian 3-geometries, as can be seen by comparing the global
form (\ref{glob}) with the local form (\ref{1.6a}). The product of
two global integrals in (\ref{glob}) cannot be as sensitive as the
local form. Indeed, it seems to us that the twin principles of
best matching and the local square root may implement the
Cartesian ideal of explaining all dynamics by geometry.

\begin{acknowledgements} We thank Domenico (Nico) Giulini for his
constructive criticism and helpful insights, especially in
connection with the 3+1 decomposition for the massive vector
field. We thank Bruno Bertotti for pointing out the importance of
establishing the status of Brans--Dicke theory within our
approach.
Their input is largely responsible for the main advances in this
paper as compared with its original draft. Comments of an
unkown referee on our original submission have also led, we trust, to
a significantly clearer presentation. We also thank Edward Anderson
for helpful comments. Finally, our title is an obvious tribute to John
Wheeler.
\end{acknowledgements}

\end{document}